\documentclass[sigconf,nonacm]{acmart}
\usepackage{comment}    
\usepackage[disable]{todonotes} 
\newcommand{\BGI}[1]{\todo[color=shadered,inline,size=footnotesize]{\textbf{Boris says:$\,$} #1}} 
\newcommand{\BG}[1]{\todo[color=shadered,inline,size=tiny,fancyline]{\textbf{Boris says:$\,$} #1}} 

\usepackage{xspace}     
\usepackage{etoolbox}   

\usepackage{amsmath}    
\usepackage{mathtools}  
\usepackage{stmaryrd}   

\usepackage{graphicx}   
\usepackage{framed}

\usepackage{listings}    

\usepackage{caption}
\usepackage{subcaption}  
\usepackage{colortbl}
\usepackage{multirow}

\usepackage{cleveref}    

\usepackage{hhline}
\usepackage{adjustbox}
\newcounter{todocounter}
\usepackage{algorithm}
\usepackage[noend]{algpseudocode}
\usepackage{empheq}
\usepackage{fancyvrb}
\usepackage{enumitem}
\VerbatimFootnotes

\newcommand{\pbds}{PBDS\xspace}
\newcommand{\abbrAQP}{AQP\xspace}
\newcommand{\abbrPBDS}{PBDS\xspace}
\newcommand{\qhighcrimes}{\ensuremath{Q_{highcrime}}\xspace}

\newcommand{\card}[1]{\ensuremath{\mid #1 \mid}}
\newcommand{\absolute}[1]{\ensuremath{\card{#1}}}
\newcommand{\expectation}[2]{\ensuremath{\mathbb{E}_{\sim{} {#2}}[#1]}\xspace}

\newcommand{\relSchema}{{\bf R}}
\newcommand{\schemaOf}[1]{\textsc{SCH}({#1})}

\newcommand{\att}{\ensuremath{a}\xspace}
\newcommand{\rel}{\ensuremath{R}\xspace}
\newcommand{\db}{\ensuremath{D}\xspace}
\newcommand{\query}{\ensuremath{Q}\xspace}
\newcommand{\tup}{t}

\newcommand{\attset}{A}

\newcommand{\parti}{F}
\newcommand{\rparti}{F}

\newcommand{\domain}[1]{\mathcal{D}(#1)}
\newcommand{\range}{\ensuremath{r}\xspace}
\newcommand{\ranges}{\ensuremath{\mathcal{R}}\xspace}
\newcommand{\numranges}{\ensuremath{n(\mathcal{R})}\xspace}
\newcommand{\group}[1]{\mathcal{G}(#1)}
\newcommand{\satgroup}[1]{\mathcal{G'}(#1)}

\newcommand{\prov}[2]{P(#1,#2)}

\newcommand{\provSketch}{\ensuremath{\mathcal{P}}}
\newcommand{\provranges}[3]{\ranges(#3,#1,#2)}
\newcommand{\instOf}[1]{\db_{#1}}

\newcommand{\schema}[1]{\textsc{Sch}(#1)}

\newcommand{\relInst}[1]{\ensuremath{\rel_{#1}}\xspace}
\newcommand{\psSet}{\ensuremath{\mathcal{PS}}\xspace}

\newcommand{\safeatt}{\ensuremath{\text{\textsc{safe}}}\xspace}

\newcommand{\candA}{\ensuremath{\attset_{cand}}\xspace}
\newcommand{\sizename}{\ensuremath{size}\xspace}

\newcommand{\esize}[5]{\ensuremath{\widetilde{size}(#1,#2,#3,#4,#5)}\xspace}
\newcommand{\asize}[5]{\ensuremath{size(#1,#2,#3,#4,#5)}\xspace}

\newcommand{\aesize}{\esize{\query}{\db}{\rel}{\att}{\ranges}\xspace}
\newcommand{\aasize}{\asize{\query}{\db}{\rel}{\att}{\ranges}\xspace}

\newcommand{\rse}{\ensuremath{\text{\textsc{RSE}}\xspace}\xspace}
\newcommand{\sel}{\ensuremath{\text{\textsc{Sel}}\xspace}\xspace}
\newcommand{\esel}{\ensuremath{\widetilde{\text{\textsc{Sel}}\xspace}}\xspace}
\newcommand{\aqr}{\tilde{Q}(S)\xspace}
\newcommand{\samples}[2]{\ensuremath{S(#1,#2)}}
\newcommand{\samplerate}{\theta}

\newcommand{\GID}{GID}
\newcommand{\pred}[1]{\ensuremath{Pr(#1)}}
\newcommand{\satps}[2]{\ensuremath{f(#1,#2)}}
\newcommand{\satranges}{\mathcal{R}_{sat}}
\newcommand{\predicate}{Predicate(t)\xspace}

\newcommand{\dsCrime}{CRIME\xspace}
\newcommand{\dsTPCH}{TPC-H\xspace}
\newcommand{\dsParking}{PARKING\xspace}
\newcommand{\dsStars}{STARS\xspace}

\newcommand{\cthead}{\cellcolor{tableHeadGray}}
\colorlet{LightBlue}{blue!30!}
\colorlet{LightRed}{red!20!}
\colorlet{LightGreen}{green!30!}

\colorlet{LRed}{red!60!}
\colorlet{LGreen}{green!70!}
\colorlet{LBlue}{blue!70!}

\definecolor{white}{rgb}{1,1,1}
\definecolor{black}{rgb}{0,0,0}
\definecolor{lgrey}{rgb}{0.9,0.9,0.9}
\definecolor{llgrey}{rgb}{0.93,0.93,0.93}
\definecolor{lllgrey}{rgb}{0.96,0.96,0.96}
\definecolor{grey}{rgb}{0.7,0.7,0.7}
\definecolor{dgrey}{rgb}{0.5,0.5,0.5}
\definecolor{red}{rgb}{1,0,0}
\definecolor{green}{rgb}{0,1,0}
\definecolor{yellow}{rgb}{1.0, 1.0, 0.0}
\definecolor{darkgreen}{rgb}{0,0.5,0}
\definecolor{darkblue}{rgb}{0,0,0.5}
\definecolor{darkpurple}{rgb}{0.5,0,0.5}
\definecolor{darkdarkpurple}{rgb}{0.3,0,0.3}
\definecolor{blue}{rgb}{0,0,1}
\definecolor{shadegreen}{rgb}{0.95,1,0.95}
\definecolor{shadeblue}{rgb}{0.95,0.95,1}
\definecolor{shadered}{rgb}{1,0.85,0.85}
\definecolor{oddRowGrey}{rgb}{0.95,0.95,0.95}
\definecolor{evenRowGrey}{rgb}{0.85,0.85,0.85}
\definecolor{tableHeadGray}{rgb}{0.85,0.85,0.85}
\definecolor{lightgrey}{rgb}{0.88,0.88,0.88}
\definecolor{lightyellow}{rgb}{1.0, 1.0, 0.88}
\definecolor{selectiveyellow}{rgb}{1.0, 0.73, 0.0}

\numberwithin{theo}{section}

\numberwithin{exam}{section}

\numberwithin{defi}{section}

\newtheorem{Definition}{Definition}

\newtheorem{Example}{Example}

\newcommand{\parttitle}[1]{\smallskip\noindent\textbf{#1.}}

\newcommand{\projection}{\ensuremath{\Pi}}
\newcommand{\selection}{\ensuremath{\sigma}}
\newcommand{\aggregation}{\ensuremath{\gamma}}
\newcommand{\Aggregation}[2]{{}_{#1}\aggregation_{#2}}
\newcommand{\union}{\ensuremath{\cup}}
\newcommand{\intersection}{\ensuremath{\cap}}
\newcommand{\difference}{\ensuremath{-}}

\newcommand{\join}{\ensuremath{\bowtie}}
\newcommand{\crossprod}{\ensuremath{\times}}
\newcommand{\duprem}{\ensuremath{\delta}}
\newcommand{\win}{\ensuremath{\omega}}

\newcommand{\bag}[1]{\{\!\!\{ {#1} \}\!\!\}}

\newcommand{\stratRANall}{\textbf{RAND-ALL}\xspace}
\newcommand{\stratRANrelall}{\textbf{RAND-REL-ALL}\xspace}
\newcommand{\stratRANgb}{\textbf{RAND-GB}\xspace}
\newcommand{\stratRANpk}{\textbf{RAND-PK}\xspace}
\newcommand{\stratRANagg}{\textbf{RAND-AGG}\xspace}

\newcommand{\stratCBOPTrel}{\textbf{CB-OPT-REL}\xspace}
\newcommand{\stratCBOPTgb}{\textbf{CB-OPT-GB}\xspace}
\newcommand{\stratCBOPT}{\textbf{CB-OPT}\xspace}
\newcommand{\stratOPT}{\textbf{OPT}\xspace}
\newcommand{\stratNoPS}{\textbf{NO-PS}\xspace}
\newcommand{\QAGH}{\textbf{Q-AGH}\xspace}
\newcommand{\QAJGH}{\textbf{Q-AJGH}\xspace}
\newcommand{\QAAGH}{\textbf{Q-QAAGH}\xspace}
\newcommand{\QAAJGH}{\textbf{Q-QAAJGH}\xspace}

\usepackage{hyperref}
\usepackage{cleveref}
\usepackage{xspace}

\crefname{Example}{ex.}{ex.}
\Crefname{Example}{Ex.}{Ex.}
\Crefname{figure}{Fig.}{Fig.}
\Crefname{section}{Sec.}{Sec.}
\Crefname{Definition}{Def.}{Def.}
\Crefname{Theorem}{Thm.}{Thm.}
\Crefname{Lemma}{Lem.}{Lem.}

\newcommand{\mathtext}[1]{\thickspace\text{#1}\thickspace}

{
\noindent \textsc{Proof Sketch.}%
}%
{\qedsymbol}

\begin{document}

\graphicspath{ {figs/} }

\definecolor{lstpurple}{rgb}{0.5,0,0.5}
\definecolor{lstred}{rgb}{1,0,0}
\definecolor{lstreddark}{rgb}{0.7,0,0}
\definecolor{lstredl}{rgb}{0.64,0.08,0.08}
\definecolor{lstmildblue}{rgb}{0.66,0.72,0.78}
\definecolor{lstblue}{rgb}{0,0,1}
\definecolor{lstmildgreen}{rgb}{0.42,0.53,0.39}
\definecolor{lstgreen}{rgb}{0,0.5,0}
\definecolor{lstorangedark}{rgb}{0.6,0.3,0}
\definecolor{lstorange}{rgb}{0.75,0.52,0.005}
\definecolor{lstorangelight}{rgb}{0.89,0.81,0.67}
\definecolor{lstbeige}{rgb}{0.90,0.86,0.45}

\DeclareFontShape{OT1}{cmtt}{bx}{n}{<5><6><7><8><9><10><10.95><12><14.4><17.28><20.74><24.88>cmttb10}{}

\lstdefinestyle{psql}
{
tabsize=2,
basicstyle=\small\upshape\ttfamily,
language=SQL,
morekeywords={PROVENANCE,BASERELATION,INFLUENCE,COPY,ON,TRANSPROV,TRANSSQL,TRANSXML,CONTRIBUTION,COMPLETE,TRANSITIVE,NONTRANSITIVE,EXPLAIN,SQLTEXT,GRAPH,IS,ANNOT,THIS,XSLT,MAPPROV,cxpath,OF,TRANSACTION,SERIALIZABLE,COMMITTED,INSERT,INTO,WITH,SCN,UPDATED},
extendedchars=false,
keywordstyle=\bfseries,
mathescape=true,
escapechar=@,
sensitive=true
}

\lstdefinestyle{psqlcolor}
{
tabsize=2,
basicstyle=\small\upshape\ttfamily,
language=SQL,
morekeywords={PROVENANCE,BASERELATION,INFLUENCE,COPY,ON,TRANSPROV,TRANSSQL,TRANSXML,CONTRIBUTION,COMPLETE,TRANSITIVE,NONTRANSITIVE,EXPLAIN,SQLTEXT,GRAPH,IS,ANNOT,THIS,XSLT,MAPPROV,cxpath,OF,TRANSACTION,SERIALIZABLE,COMMITTED,INSERT,INTO,WITH,SCN,UPDATED},
extendedchars=false,
keywordstyle=\bfseries\color{lstpurple},
deletekeywords={count,min,max,avg,sum},
keywords=[2]{count,min,max,avg,sum},
keywordstyle=[2]\color{lstblue},
stringstyle=\color{lstreddark},
commentstyle=\color{lstgreen},
mathescape=true,
escapechar=@,
sensitive=true
}

\lstdefinestyle{datalog}
{
basicstyle=\footnotesize\upshape\ttfamily,
language=prolog
}

\lstdefinestyle{pseudocode}
{
  tabsize=3,
  basicstyle=\small,
  language=c,
  morekeywords={if,else,foreach,case,return,in,or},
  extendedchars=true,
  mathescape=true,
  literate={:=}{{$\gets$}}1 {<=}{{$\leq$}}1 {!=}{{$\neq$}}1 {append}{{$\listconcat$}}1 {calP}{{$\cal P$}}{2},
  keywordstyle=\color{lstpurple},
  escapechar=&,
  numbers=left,
  numberstyle=\color{lstgreen}\small\bfseries,
  stepnumber=1,
  numbersep=5pt,
}

\lstdefinestyle{xmlstyle}
{
  tabsize=3,
  basicstyle=\small\upshape\ttfamily,
  language=xml,
  extendedchars=true,
  mathescape=true,
  escapechar=£,
  tagstyle=\bfseries,
  usekeywordsintag=true,
  morekeywords={alias,name,id},
  keywordstyle=\color{lstred}
}

\lstdefinestyle{xmlstyle-color}
{
  tabsize=3,
  basicstyle=\small\upshape\ttfamily,
  language=xml,
  extendedchars=true,
  mathescape=true,
  escapechar=£,
  tagstyle=\color{keywordpurple},
  usekeywordsintag=true,
  morekeywords={alias,name,id},
  keywordstyle=\color{lstred}
}

\lstset{style=psqlcolor}

\title{Cost-based Selection of Provenance Sketches for Data Skipping}

\author{Ziyu Liu}
\affiliation{\institution{Illinois Institute of Technology}}
\email{zliu@hawk.iit.edu}

\author{Boris Glavic}
\affiliation{\institution{University of Illinois, Chicago}}
\email{bglavic@uic.edu}

\begin{abstract}
  Provenance sketches, light-weight indexes that record what data is needed (\emph{is relevant}) for  answering a query, can significantly improve performance of important classes of queries (e.g., \lstinline!HAVING! and top-k queries). Given a horizontal partition of a table, a provenance sketch for a query $Q$ records which fragments contain provenance. Once a provenance sketch has been captured for a query, it can be used to speed-up subsequent queries by skipping data that does not belong to a sketch. The size and, thus, also the effectiveness of a provenance sketch is often quite sensitive to the choice of attribute(s) we are partitioning on. In this work, we develop sample-based estimation techniques for the size of provenance sketches akin to a specialized form of approximate query processing. This technique enables the online selection of provenance sketches by estimating the size of sketches for a set of candidate attributes and then creating the sketch that is estimated to yield the largest benefit. We demonstrate experimentally that our estimation is accurate enough to select optimal or near optimal provenance sketches in most cases which in turn leads to a runtime improvement of up to \%60 compared to other strategies for selecting provenance sketches.
\end{abstract}

\maketitle


\section{Introduction}
\label{sec:introduction}
For many decades, database systems have used physical design techniques such as index structures~\cite{guttman1984r}, zone maps, and horizontal partition to provide fast access to data. Given a query, the DBMS determines statically (before the query is run) whether any physical design artifacts can be utilized to reduce the amount of data that needs to be accessed to answer the query. However, for important types of queries such as top-k and \lstinline!HAVING! queries, this static analysis is insufficient for determining what data is relevant. As we have demonstrated in~\cite{ps2021}, we can instead determine dynamically (at query runtime) which data is relevant by using compact over-approximations of which subset of the database is relevant using a data structure called \emph{provenance sketch}. Given a horizontal range-partitioning of a table (that does not have to correspond to the physical data layout of the table), a provenance sketch records what fragments of the partition contain at least one row of provenance. In~\cite{ps2021}, we introduced techniques for generating sketches, for determining when a sketch created for a query $Q_1$ can be used to answer a query $Q_2$, and for answering queries with sketches by translating sketches into selection conditions that filter out data that does not belong to the sketch. The DBMS can then utilize existing physical design artifacts to efficiently evaluate such selection conditions. Together we refer to this suite of techniques as \emph{provenance-based data skipping} or \emph{\pbds} in short. \pbds can significantly improve the runtime of queries by unearthing opportunities for the DBMS to exploit physical design for queries where this was previously not possible.

In the online version of \pbds, a system has to decide for each incoming query whether to create a new sketch for the query and pay the overhead of sketch capture, whether to utilize an existing sketch to answer the query which would improve the query's runtime, or just to evaluate the query without using a sketch. When creating a new sketch, an important choice is what horizontal range-partition to use (on which attribute and using which ranges) as the choice of attribute (and ranges) can significantly impact the amount of data covered by the sketch (its ``size'') and, thus, its effectiveness for data skipping.
In this work, we develop efficient techniques for estimating the size of sketches upfront to be able to select the partition to build a sketch on in a principled manner using novel approximate query processing (AQP) techniques.

\begin{figure*}[t]
  \begin{minipage}{0.39\textwidth}
    \centering
  \begin{subfigure}{1\linewidth}
    \centering
    \begin{minipage}{0.9 \textwidth}
      \captionsetup{singlelinecheck = false, justification=justified}
        \lstset{upquote=true,frame=single, title = \bf{\qhighcrimes}}
        \begin{lstlisting}
SELECT sum(numcrimes) AS totcrimes,
       pid, month, year
FROM crimes
GROUP BY pid,month,year
HAVING num_crimes >= 100;
\end{lstlisting}
    \end{minipage}    
    \caption{Running example query}
        \label{fig:q1}
    \end{subfigure}
          \begin{subfigure}{1\linewidth}
        \centering
        \begin{tabular}{|c|c|c|c|c} \hhline{|-|-|-|-|~}
          \cthead totcrimes & \cthead pid & \cthead month & \cthead year \\
          \hhline{|-|-|-|-|~}
        184                 & 4           & 1             & 2013         \\
        192                 & 8           & 6             & 2015         \\
        157                 & 2           & 7             & 2016         \\
		\hhline{|-|-|-|-|~}
        \end{tabular}
        \caption{Result of \qhighcrimes}
        \label{fig:results-of-Q-1}
    \end{subfigure}
  \end{minipage}
  \begin{minipage}{0.59 \textwidth}
    \centering 
  \begin{subfigure}{1\linewidth}
\newcommand{\psisin}[1]{\fcolorbox{black}{black}{\color{white}\ensuremath{\mathbf{#1}}}}
\newcommand{\psisout}[1]{\ensuremath{\textcolor{gray}{\mathbf{#1}}}}

      \begin{tabular}{c|c|c|c|c|l|c|c|c|}
        \hhline{~|-|-|-|-|~|-|-|-|}
        & \cthead pid & \cthead month & \cthead year  & \cthead numcrimes &  & \textbf{$\provSketch_{pid}$}                           & \textbf{$\provSketch_{month}$}                       & \textbf{$\provSketch_{year}$}                         \\
        \hhline{~|-|-|-|-|~|-|-|-|}
       	$g_0$   & 3           & 1             & 2010          & 88                 &  & \cellcolor{shadeblue}  \psisin{p_1}                    & \cellcolor{shadeblue}                                & \cellcolor{shadeblue} \multirow[]{-1}*{\psisout{y_1}} \\
        \cline{2-5}
         $g_1$  & \textbf{4}  & \textbf{1}    & \textbf{2013} & \textbf{73}        &  & \cellcolor{shadegreen}                                 & \cellcolor{shadeblue} \multirow[]{-1}*{\psisin{m_1}} & \cellcolor{shadegreen}                                \\
                & \textbf{4}  & \textbf{1}    & \textbf{2013} & \textbf{101}       &  & \cellcolor{shadegreen} \multirow[]{-2}*{\psisin{p_2}}  & \cellcolor{shadeblue}                                & \cellcolor{shadegreen}                                \\
               \cline{2-5}
        $g_2$   & \textbf{8}  & \textbf{6}    & \textbf{2015} & \textbf{86}        &  & \cellcolor{shadered}                                   & \cellcolor{shadegreen}                               & \multirow[]{-1}*{\psisin{y_2}}\cellcolor{shadegreen}  \\
                & \textbf{8}  & \textbf{6}    & \textbf{2015} & \textbf{96}        &  & \cellcolor{shadered} \multirow[]{-2}*{\psisin{p_3}}    & \cellcolor{shadegreen}\multirow[]{-1}*{\psisin{m_2}} & \cellcolor{shadegreen}                                \\
        \cline{2-5}
         $g_3$  & \textbf{2}  & \textbf{7}    & \textbf{2016} & \textbf{157}       &  & \cellcolor{shadeblue}   \multirow[]{-1}*{\psisin{p_1}} & \cellcolor{shadegreen}                               & \cellcolor{shadegreen}                                \\
        \cline{2-5}
          $g_4$ & 7           & 2             & 2022          & 83                 &  & \cellcolor{shadered}                                   & \cellcolor{shadeblue} \psisin{m_1}                   & \cellcolor{shadered}                                  \\
          \cline{2-5}
        $g_5$   & 7           & 9             & 2023          & 58                 &  & \cellcolor{shadered} \multirow[]{-2}*{\psisin{p_3}}    & \cellcolor{shadered}\multirow[]{-1}*{\psisout{m_3}}  & \cellcolor{shadered}\multirow[]{-2}*{\psisout{y_3}}   \\
                & \ldots      & \ldots        & \ldots        & \ldots             &  & \ldots                                                 & \ldots                                               & \ldots                                                \\
       \hhline{~|-|-|-|-|~|-|-|-|}
        \end{tabular}
        \caption{Example Crime Statistics Table and three range partitions (on attributes \texttt{pid}, \texttt{month}, and \texttt{year}). Fragments belonging to the provenance sketch for  are shown with black background.}
        \label{fig:CRIMES}
      \end{subfigure}

  \end{minipage}

    \caption{Running example illustrating the importance of choosing the right attribute to build a provenance sketches on.}
    \label{fig:running-example-with-sketches}
\end{figure*}

\begin{Example}\label{ex-ps}
We illustrate the need for selecting an appropriate partition when creating sketches by means of an example.
\Cref{fig:q1} shows a query which returns the total number of recorded crimes grouped by police district (identified by an pid), month, and year for groups that have more than 100 recorded crimes. \Cref{fig:CRIMES} shows an example instance of the crime dataset. Rows belonging to the provenance of the query are highlighted in bold.
\footnote{For certain queries, not all choices of range-partitioning are safe in that the generated sketches may not produce the same query result. The safety test from~\cite{ps2021} can be applied to test statically (at query compile time) which sets of attributes are safe for creating sketches. We do not elaborate on this further, but will just pre-filter sketch candidates using this test.}
Let us compare three sketches for this query, one created based on partitioning the crimes data on attribute \texttt{pid}, the second one for attribute \texttt{month} and the third one for attribute \texttt{year}.
  Let us assume that for attributes \texttt{pid},\texttt{month} and \texttt{year} we use a range partitioning with ranges $\ranges_{pid}$, $\ranges_{month}$ and $\ranges_{year}$  that are shown below. For instance, such ranges can be extracted from equi-depth histograms that databases maintain as statistics for result size estimation during query optimization~\cite{ps2021}. 
 \footnote{In realistic settings, PBDS typically uses partitions with 100s or 1000s of fragments. 
 }
  \[
    \ranges_{pid} = \{ \range_1 = [1,3], \range_2 = [4,6], \range_3 = [7,9] \}
  \]
  \[
    \ranges_{month} = \{ \range_1' = [1,4], \range_2' = [5,8], \range_3' = [9,12] \}
  \]
  \[
    \ranges_{year} = \{ \range_1'' = [2010,2012], \range_2'' = [2013,2020], \range_3'' = [2021,2024] \}
  \]
Let us use $p_i$ to denote the fragment for range $\range_i \in \ranges_{pid}$, $m_i$ to denote the fragment for range $\range_i' \in \ranges_{month}$ and $y_i$ to denote the fragment for range $\range_i'' \in \ranges_{year}$. Using colored background, \Cref{fig:CRIMES} shows which fragments each row belongs too for the three candidate parsons.
A provenance sketch based on a range partition consists of all ranges whose fragments contain one or more rows of provenance. The names of such fragments are shown with black background in \Cref{fig:CRIMES}.
A sketch can be compactly encoded as this set of ranges, or, factoring out the ranges, as a bit vector that contains one bit per range, which is $1$ if the range is included in the sketch. For example, the sketch build on $\ranges_{pid}$ contains all three ranges. That is, the data encoded by the sketch is the full input table. In contrast, a sketch on $\ranges_{year}$ is much more effective, containing only a single range $\range_2''$ whose fragment contains precisely the rows belonging to the provenance of the query. Thus, for the given query and database, the optimal choice for building a sketch would be using a range partition on attribute \texttt{year} with ranges $\ranges_{year}$. Once a sketch has been created, it can be used to evaluate future executions of our running example query and of similar queries (techniques presented in~\cite{ps2021} can be used to determine with a sketch captured for a query $\query_1$ can be used to answer a query $\query_2$) by filtering out data that does not belong to the sketch. This is achieved by adding a disjunction of range conditions for all ranges belonging to the sketch. For instance, for the sketch build on \texttt{year}, we would use the following condition:
\begin{lstlisting}
WHERE year BETWEEN 2013 AND 2020
\end{lstlisting}

\end{Example}

Intuitively, the difference in sketch sizes stem from how ``aligned'' a partition is with the provenance of a query.
\BG{Q-1 in the running example is different from the one in \Cref{runtime_q1}. Use a different name for the example query?}
As an anecdotal result demonstrating the benefits of sketch selection, consider \Cref{runtime_q1} which shows the runtime of a query similar to our running example (differing only in the group-by attributes and having conditions) evaluates over the crime dataset from \url{crimes}\BG{@Ziyu turn this into a reference in the bibliography}. Without the use of provenance sketches (No-PS) the query takes $\sim 10$ seconds. Even a poor choice of sketch (on attribute \texttt{numcrimes}) improves performance by a factor of 2. However, an optimal choice (attribute \texttt{district}) reduces runtime to 2 seconds (a factor of 5).

In this work, we study techniques that enable a system to choose effective provenance sketches. We start by introducing a size estimation technique for sketches which given a query $\query$, attribute(s) $A$ on which to build a sketch on, and ranges $\ranges_A$, estimates the size of a sketch building on $A$ using $\ranges_A$ using sample-based techniques rooted in approximate query processing (\emph{AQP}). Ignoring physical design (e.g., an index may be available for filtering data based on a sketch on a particular attribute), the size of a sketch, i.e., the fraction of a table's data covered by a sketch is a good predictor of the effectiveness of the sketch for improving query performance. We develop estimators for the number of ranges a sketch will contain that are based on stratified sampling and bootstrap. One challenge with this is that we essentially have to estimate a distinct count: the number of distinct fragments that are estimated to contain provenance.\BG{@Ziyu: can we say something smart / more detailed about how our techniques does that than what I am writing above?}
We identify conditions under which the samples created for one query can be reused to estimate sketch sizes for other queries. We implemented the sampling techniques in Postgres and integrate size estimation of sketches into the prototype implementation of PBDS from~\cite{ps2021}.

We then discuss several strategies for selecting what attribute to build a sketch on ranging from strategies that just statically analyze the query to determine which attributes are ``important'' for the query (e.g., attributes not accessed by the query are typically not good candidates for building sketches) to strategies that estimate the size of every possible set of candidate attributes and choose the one with the lowest estimated size. In our experimental evaluation we evaluate the quality of the size estimates produced by our techniques and compare the aforementioned strategies. We demonstrate that capturing provenance sketches on group-by attributes, but using our size estimation techniques to select an approximately optimal subset of group-by attributes.


 \parttitle{Contributions}
Our main technical contributions are:

\begin{itemize}[noitemsep,topsep=0pt,parsep=0pt,partopsep=0pt,leftmargin=*]
\item We techniques for estimating the size of provenance sketches based on samples. These techniques includes stratified sampling, aggregations estimators are based in approximate query processing, but estimate the amount of distinct ranges belonging to a sketch rather than the query result. \BG{@Ziyu: please say something about the techniques used in here: sampling strategies, estimators that we reuse, ...}
\BG{@Ziyu: please state something about our formal results, i.e., what do we prove?}
\item We investigate several strategies for selecting provenance sketches that differ in whether the are based on sketch size estimation (we call such strategies \emph{cost-based}) or randomized choice and on whether/how they prune candidates using static analysis of the input query to determine which attributes may be ``relevant'' for the query. \textbf{we prove that the GB-cost-based is the optimal strategy where generate size estimation on groupby attributes canidates for the most query cases in our supported query templates.}
\item We implement our size estimation techniques on top of Postgres in the PBDS prototype presented in~\cite{ps2021}. We demonstrate experimentally that our size estimation techniques have decent accuracy and acceptable overhead. Specifically, the accuracy is sufficient for selecting the optimal sketch (in term of size) in nearly 100\% of all tested scenarios.
Regarding the different strategies, while the cost-based strategy that estimates the size of sketches for all possible candidate attributes typically yields the smallest sketch, the optimal choice is typically to make a cost-based choice over a pruned set of attributes that is restricted to attributes relevant to the query as pruning the number of candidates reduces the overhead of size estimation.
\end{itemize}

\BGI{Update the last paragraph once we are completely happy with the structure}
The remainder of this paper is organized as follows. We cover background in \Cref{sec:bcakground} and review related work in \Cref{sec:related-work}. Afterwards, we define provenance sketches in \Cref{sec:provenance-sketch}. Our definition and introduction of cost model is discussed in \Cref{sec:cost-model}. In \Cref{sec:rules}, we present techniques for determining which candidate attribute is best for capturing provenance sketch. We present experimental results in \Cref{sec:experiments} and conclude in \Cref{sec:concl-future-work}.
\begin{table}
    \centering
    \begin{tabular}{|c|c|c|c|c|} \hhline{|-|-|-|-|-|}
    \cthead &\cthead No-PS & \cthead district & \cthead zipcode   & \cthead numcrimes \\
    \qhighcrimes &10.148 & 1.973  & 2.051 & 5.092 \\
    \hhline{|-|-|-|-|-|}
    \end{tabular}
    \caption{Running time (sec) of \qhighcrimes without sketches (No-PS) versus using sketches on specific attributes.}
    \label{runtime_q1}
\end{table}


\section{Background}
\label{sec:bcakground}

\subsection{The Relational Data Model and Algebra}\label{sec:rel-algebra}

A relation schema $R(a_1,...,a_n)$ consists of a name ($R$) and a list of attribute names. The arity of the schema is the number of attributes in the schema. The table below gives the definition of the bag semantics version of relational algebra. $\schema{\query}$ is used to denote the schema of the result of query \query. $\query(\db)$ is used to denote the result of \query over database \db. We use $t^{n} \in R$ to denote that the multiplicity of tuple $t$ in relation $R$ is  $n$. Selection $\selection_\theta(R)$ returns all the tuples from R which satisfy condition $\theta$.   Projection $\projection_A(R)$ returns all the input tuples on a list of projection expressions. Union $R \union S $ returns the union of all the tuples in relation R and relation S. Intersection $R \intersection S$ returns the tuples which both exist in relation R and relation S. Difference $R \difference S$ returns the tuples which are in relation R but not in relation S.
  Cross product $R \crossprod S$ returns the tuples which is the product of each tuple in relation R and each tuple relation S. Aggregation $G\gamma_{f(a)}(R)$ groups tuples according to their value  in attribute $G$ then computes the aggregation function $f$ over the values of attribute $a$ for each group. Duplicate removal $\duprem (R)$ remove the duplicate tuples in relation R. The window operator $\win_{f(a)\rightarrow x, G\parallel O}(R)$ returns tuples with additional attribute $x$ whose value is the result of aggregation function $f$.
  Aggregation function $f$ is applied on the window generated by partitioning the input on $G \subseteq \schema{R}$ and including only tuples which are smaller than t wrt. their values in attributes $O \subseteq \schema{R}$ where $G \intersection O = \emptyset$. \\

 \begin{figure}[h]
 \begin{tabular}{|c|l|}
 \hline \rowcolor{lightgrey}
   Operator & Definition\\
 \hline
  $\selection$& $\selection_\theta(R) = \{t^n \mid t^n \in R \wedge t = \theta\}$\\
  \hline
  $\projection$& $\projection_A(R) = \{t^n \mid n = \Sigma_{u.A = t}R(u)\}$\\
  \hline
  $\union$& $R \union S= \{t^{n+m} \mid t^n \in R \wedge t^m \in S\}$\\
 \hline
  $\intersection$& $R \intersection S= \{t^{min(n,m)} \mid t^n \in R \wedge t^m \in S\}$\\
 \hline
 $\difference$& $R \difference S= \{t^{max(n-m),0} \mid t^n \in R \wedge t^m \in S\}$\\
 \hline
  $\crossprod$& $R \crossprod S= \{(t,s)^{n*m} \mid t^n \in R \wedge s^m \in S\}$\\
  \hline
  $\gamma$& $\Aggregation{f(a)}{G}(R) = \{(t.G, f(G_t))^1 \mid t \in R \}$\\
          & $G_t=\{(t_1.a)^n \mid {t_1}^n \in R \wedge t_1.G = t.G\}$\\
  \hline
  $\duprem$& $\duprem (R)= \{t^1\mid t\in R\}$\\
  \hline
  $\win$& $\win_{f(a)\rightarrow x, G\parallel O}(R) \equiv \{(t, f(P_t))^n \mid t^n \in R \}$\\
          & $P_t=\{(t_1.a)^n \mid {t_1}^n \in R \wedge t_1.G = t.G \wedge t_1\leq O^t\}$\\
        \hline
 \end{tabular}
\BG{we can see whether we need to keep all operators if we do not use all}
 \caption{Relational Algebra}
\label{fig:ex-relational-algebra}
\end{figure}


\subsection{Provenance and Sufficiency}\label{sec:query-provenance}

In the following, we are interested in finding subsets $\db'$ of an input database $\db$ that are sufficient for answering a query $\query$. That is, for which $\query(\db') = \query(\db)$. We refer to such subsets as sufficient inputs.

\begin{Definition}[Sufficient Input]\label{def:sufficient}
Given a query $\query$ and database $\db$, we call $\db' \subseteq \db$ \emph{sufficient} for $\query$ wrt. $\db$ if \\
	$$Q(D) = Q(D')$$.
\end{Definition}

Several provenance models for relational queries have been proposed in the literature~\cite{CC09}. Most of these models have been proven to be instances of the semiring provenance model~\cite{GK07,GT17} and its extensions for difference/negation~\cite{GP10} and aggregation~\cite{AD11d}. Our main interest in provenance is to determine a sufficient subset of the input database. Thus, even a simple model like Lineage~\cite{G21} which encodes provenance as a subset of the input database is expressive enough. We use $\prov{\query}{\db}$ to denote the provenance of a query $\query$ over database $\db$ encoded as a bag of tuples and assume that $\prov{\query}{\db}$ is sufficient for $\query$ wrt. $\db$. For instance, we may construct $\prov{\query}{\db}$ as the union of the Lineage for all  tuples  $\tup \in \query(\db)$.
Our results hold for any provenance model that guarantees sufficiency.

\subsection{Stratified Sampling}\label{sec:background-stratified-sample}
\BG{It is not clear why we are talking about stratified sampling and histograms not being suitable for what?}
\textbf{Histogram is a good choice for approximating simple aggregation results such as aggregation query with 1 group by. It can provide reliable approximating results. However, }
The histogram is not suitable for the complicated queries such as the selective aggregation queries with  more than 3 group by attributes for the reason that the estimation of cost using higher-dimensions histogram is not accurate under uniformly assumption considering the input data could be skewed or normal distributed. \textbf{What's more, the correlation of each attribute could also affect the accuacy of estimation based on the historgram.} Sampling-based approximating processing could be a better choice. The simplest such data structure is a uniform sample. Considering the numerical data P, from P, we can first sample a subset S of size K uniformly. Every tuple is sampled with equal probability. However, the pure uniform sampling will suffer from the highly selective queries not containing the relevant data. And this can result confusing and misleading results. Stratified sampling is one way to mitigate the effects of selective predicates which prioritizes sampling certain regions of the database. As a result, drafting the samples by stratified sampling technique can be more effective than the traditional uniformly sampling.
 Instead of directly sampling from P, we first partition P into B strata, which are mutually exclusive partitions defined by groupings over attribute C in P. Within each stratum $P_1,...P_B$, we uniformly sample resulting in samples $S_1,...S_B$. So, instead of a single parameter K which controlled the accuracy in the uniform sampling case introduced before, we have a $K_1,...K_B$ for each stratum

\subsection{Selectivity Estimation}
\BG{It's not clear at this point why we are talking about selectivity estimation. We need to explain how this relates to your work, otherwise why do we need to learn about it at this point in time.}
For each query, there are many equivalent execution plans. To choose the most efficient among these different query plans, the optimizer needs to estimate their cost. The cost is mainly decided by the selectivity which means the percentage of rows selected from the input table. Computing the precise cost of each plan is usually not possible without actually evaluating the plan. Thus, instead, optimizers use statistics, such as the size of relations and the depth of the indexes and histogram to estimate the cost of each plan. For example, a selection condition consisting of a number of predicates. Each predicate has a reduction factor which is the relative reduction in the number of result tuples caused by this predicate. There are many heuristic formulas for the reduction factors of different predicates. In general, they depend on the assumption of uniform distribution of values and independence among the various relation fields. More accurate reduction factors can be achieved by maintaining more accurate statistics, for example in the form of histograms or multidimensional histograms.

\subsection{Approximate Query Processing} \label{sec:AQP}
\BG{We need to clarify why we talk about approximate query processing at this point in the paper.}
Approximate query processing (AQP) is an alternative way\BG{Alternative to what specifically?} that returns approximate answer using information which is similar to the one from which the query would be answered. It is designed primarily for aggregate queries such as count, sum and avg, etc. Given a SQL aggregate query Q, the accurate answer is $y$ while the approximate answer is $y'$. The relative error of query Q can be quantified as: $Error(Q) = \frac{y-y'}{y}$.  The goal of approximate query processing is to provide approximate answers with acceptable accuracy\BG{Clarify that this would typically be epsilon delta bounds.} in orders of magnitude less query response time than that for the exact query processing. Haas\cite{haas1997large} derived the estimators \textbf{including SUM, AVG,COUNT} and confidence intervals\BG{explain what estimators and confidence intervals are} for various aggregation functions, we utilize them to calculate the expectations of our provenance sketch candidates based on Central limit theorems, the details will be shown in \Cref{sec:Selectivity_expectation}

\section{Related Work}
\label{sec:related-work}
Our work is related to the data skipping and partition pruning~\cite{ceri1982horizontal} an essential  for data warehouses to eliminate unneeded partitions for processing the query. Different with them, our technique is based on the query provenance~\cite{tan2004research,buneman2001and}, thus provenance capture and storage are also related to our work.

\subsection{Data Skipping}
\label{sec:data-skip}
Database optimizer uses selection and projection push-down, join reordering and semijoin reducers~\cite{bernstein1981using} to avoid producing large intermediate results used for following complex operations, e.g., join, to improve the query performance. Usually these techniques through statically analyzing the query and modifying the query structure to achieve the goal of removing plenty of unneeded intermediate results.
Index is a data structure stored in the database used to quickly locate to the data  required by the query processing to minimize the number of disk accesses, e.g., $B^{+}-Tree$ index.
Data pruning is a technique which determines certain chunk of data as unnecessary for query processing and thus skipping over such data. The chunk could be rows, e.g., bitmap or bloom filters~\cite{bloom1970space}, or a set of contiguous data blocks or table partitions, i.e., partiton pruning~\cite{ceri1982horizontal}, for example,  small materialized aggregates~\cite{moerkotte1998small} , in-memory technique~\cite{lahiri2015oracle}, storage index~\cite{clarke2013storage} and hippo index~\cite{yu2016two}. Especially, when partitioning of tables represents strict clustering of data, partition pruning would significantly improve the query performance since the avoiding of unnecessary data access from the disk such that reduces the amount of physical I/O. The difference between this  technique and regular index is that data pruning is a kind of approximate index structure, which is light-weight and allows to skip large portion of data. However, it's not precise and needs to be rechecked.

\subsection{Provenance Capture and Storage}
\label{sec:prov-cap}
As mentioned in Sec.~\ref{sec:introduction} our approach uses query provenance to determine the inputs needed to answer the query. The state-of-the-art provenance capture technique~\cite{KG12,GA12} is to model provenance as annotations on data and  capture provenance by propagating annotations. For example, each output tuple t of a query Q is annotated with its provenance, i.e., a combination of input tuple annotations that explains how these inputs were used by Q to derive t.  Many database provenance systems apply this technique such as Perm~\cite{glavic2013using}, GProM~\cite{AF18}, DBNotes~\cite{bhagwat2005annotation}, LogicBlox~\cite{GA12}, Trio [29], declarative Datalog debugging~\cite{KL12}, ExSPAN~\cite{ZS10} and others. However, the annotations might be super large which are not feasible for our purpose since it is unfeasible to keep so large provenance for every query. Thus instead of annotating on each tuple, we choose to annnotate on each partition such that we only need a small space to store it. In this way we get a superset of the real provenance since each partition might also include unprovenance tuples, however, it is still a subset of the dataset.

\subsection{Approximate Query Processing using Sampling}
\BG{We need to add more citations here, look at the reference sections of papers you cite to find more papers to cite}
As we mentioned in Sec.~\ref{sec:introduction}. Our size estimation is based on the approximating query result based on the sampling technique\cite{hellerstein1997online}. In large data warehousing environments, it is often advantageous to provide fast, approximate answers to complex decision support queries using precomputed summary statistics, such as Congressional samples\cite{acharya2000congressional},stratified samples\cite{chaudhuri2007optimized,liang2021combining,kandula2016quickr}. What's more, for the queries including joins, some techniques like wander join \cite{li2016wander}, ripple join \cite{haas1999ripple} can estimate the join results efficiently. And confidence intervals can be determined\cite{haas1997large} After we have reliable estimation result, then the size of the range partitions which contained the provenance sketch can be estimated.

\subsection{Costing Queries and Cardinality Estimation}
\label{sec:cost-quer-card}
\BG{This is a big space, we need more than 2 citations here}
The Query determines the cost of executing a query plan based on the total number of rows processed at each level of a query plan, referred to as the cardinality of the plan. Cardinality estimation is a fundamental task in database query processing and optimization. Therefore, improved cardinality leads to better estimated costs and, in turn, faster execution plans. Cardinality estimation \cite{wang2020we,han2021cardinality,han2021cardinality,woltmann2019cardinality,harmouch2017cardinality,neumann2011characteristic}is derived primarily from histograms that are created when indexes or statistics are created. In our paper, after the approximating query result has been derived, we will conduct the cardinality estimation to estimate the cost of the provenance sketch, which is called cost model.

\section{Provenance Sketches and Problem Definition}
\label{sec:provenance-sketch}
As we discussed in the \Cref{sec:introduction}, we developed provenance sketches ~\cite{ps2021} before to concisely represent a superset of the provenance of a query $\query$ (a sufficient subset of the input) based on horizontal partitions of relations.
A partition is a division of a logical database or its constituent elements into distinct independent parts. The partitioning can be done by either building separate smaller databases including horizontal partitioning and vertical partitioning. There are many different criteria to split the database such as Range partitioning, Hash partitioning and List partitioning e.g. And a provenance sketch contains all fragments of the input database partitioned which contain at least one row from the $\prov{\query}{\db}$.
We limit the discussion to range-partitioning selecting a partition by determining if the partitioning key is within a certain range which will be introduced in detail in the next section, since it allows us to exploit  existing index structures when using a sketch to skip data. However, note that most of the techniques we introduced 
are independent of the type of partitioning.

  \subsection{Range Partitioning}
  \label{section:range_partition}
Given a set of intervals over the domains of a set of attributes $A \subset \relSchema$, range partitioning determines membership of tuples in fragments based on which interval their values belong to.
For simplicity, we define range partitioning for a single attribute $\att$. \Cref{fig:example-range-part} shows  two range partitions for our running example.

\begin{Definition}[Range partition] \label{def:range}
Consider a relation $\rel$ and $\att \in \relSchema$. Let $\domain{a}$ denote the domain of $\att$. Let $\ranges = \{\range_1, \ldots, \range_n\}$ be a set of intervals $[l,u] \subseteq \domain{\att}$ such that $\bigcup_{i=0}^{n} \range_i = \domain{a}$ and $r_i \cap r_j = \emptyset$ for $i \neq j$. The number of ranges in \ranges  will be denoted  $\numranges$. The \emph{range-partition} of $\rel$ on $\att$ according to $\ranges$ denoted as $\rparti_{\ranges,\att}(R)$ is defined as:
  \begin{align*}
    \parti_{\ranges,\att}(R)  = \{ \rel_{\range,\att} \mid \range \in \ranges \} \hspace{2mm}\mathtext{where}\hspace{2mm}
    \rel_{\range,\att}            = \bag{t^n \mid t^n \in \rel \wedge t.a \in  \range}
  \end{align*}
\end{Definition}

In the following we will use $\parti$ to denote a range partition if $\ranges$, $\att$ and $\rel$ are clear from the context.

\subsection{Provenance Sketch}\label{sec:back-provenance-sketches}

\BG{I am not happy with the notation $\provranges{\db}{\parti_{\ranges,a}(R)}{\query}$ here as it is quite verbose and complex. First of, the functional notation of $\parti_{\ranges,a}$ is not great in there. I think we can simplify the notation for range partition. First, use $\parti$ instead of $\parti_{\ranges,\att}(R)$, but make parameters of $\parti$ clear from the context.}

Consider a database $\db$,  query $\query$, and a range partition $\parti = \parti_{\ranges,\att}$ of $\rel$.
A provenance sketch $\provSketch$ for $\query$  according to $\parti$ is a subset of the ranges $\ranges$ of $\parti$ such that the fragments corresponding to the ranges in $\provSketch$ fully cover $\query$'s provenance within $\rel$, i.e., $\prov{\query}{\db} \cap \rel$. We use $\provranges{\db}{\parti}{\query} \subseteq \ranges$ to denote the set of ranges whose fragments contains at least one tuple from $\prov{\query}{\db}$:

\begin{align*}
  \provranges{\db}{\parti}{\query} &= \{ \range \mid \range \in \ranges \wedge \exists t \in \prov{\query}{\db}: t \in R_{\range,a} \}
\end{align*}

\begin{Definition}[Provenance Sketch]\label{def:provenance-sketch}
  Let $\query$ be a query, $\db$ a database,  $\rel$ a relation accessed by $\query$, and $\parti = \parti_{\ranges,a}(R)$ a range partition of $R$.
We call a subset $\provSketch$ of $\ranges$ a \textbf{provenance sketch} iff
$\provSketch \supseteq \provranges{\db}{\parti}{\query}$.
A sketch is called \textbf{accurate} if
$\provSketch =  \provranges{\db}{\parti}{\query}$. 
We use $\relInst{\provSketch}$, called the \textbf{instance} of $\provSketch$, to denote $\bigcup_{\range \in \provSketch} \rel_{\range,\att}$. 
\end{Definition}

Given a query $\query$ over relation $\rel$, a provenance sketch $\provSketch$ is a compact and declarative description of a superset of the provenance of $\query$ (the instance $\relInst{\provSketch}$ of $\provSketch$). We call a sketch $\provSketch$  \textit{accurate} if it only contains ranges whose fragments contain provenance. Note that we can create multiple provenance sketches for a query, each on a different table accessed by a query.
We use $\psSet$ to denote such a set of provenance sketches and use $\instOf{\psSet}$ to denote the subset of database $\db$ where all relations \rel for which there exists a sketch \provSketch in \psSet are replaced with \relInst{\provSketch}. To simplify the exposition we will focus mostly on single sketches $\psSet=\{\provSketch\}$ per query and use $\instOf{\provSketch}$ to denote $\instOf{\{\provSketch\}}$. However, our techniques are applicable for the multi-sketch case too.
Reconsider the running example shown in \Cref{fig:running-example-with-sketches} . Let $\provSketch$ be the accurate provenance sketch of $\qhighcrimes$  using the range partition on  $\ranges_{month}$. Recall that $\prov{\qhighcrimes}{\db}$ consists of all tuples from groups $g_1$, $g_2$, and $g_3$. The two fragments of the partition using $\ranges_{month}$ that contain rows from $\prov{\qhighcrimes}{\db}$ are $m_1$ and $m_2$.
Thus, an accurate provenance sketches using $\ranges_{month}$ is $\provSketch = \{ m_1, m_2 \}$.

\subsection{Sketch Safety}
We call a set of sketches \textbf{safe} for a query $\query$ and database $\db$ if evaluating $\query$ over the instance of the sketch $\instOf{\psSet}$ described by the sketches returns the same result  as evaluating it over $\db$.

\begin{Definition}[Safety]\label{def:safe}
Let $\query$ be a query and $\db$ a database.  We call a set of sketches $\psSet$ \emph{safe} for $\query$ and $\db$ iff $Q(\instOf{\psSet}) = Q(\db)$.
\end{Definition}

Obviously, only safe sketches are of interest. Following~\cite{ps2021} we define attributes to be safe for a database $\db$ and query $\query$, if any sketch created based on a range partition over these attributes is safe.

\begin{Definition}[Attribute Safety]\label{def:attribute-safety}
Let $\db$ be a database, $\query$ a query, and $\attset$ a set of attributes from the schema of a relation $\rel$ accessed by $\query$.
  We call $\attset$ \emph{safe} for  $\query$ and $\db$  
  if for every range partition $\rparti_{\ranges,\attset}$ of $\rel$, every sketch $\provSketch$ based on $\rparti_{\ranges,\attset}$ is safe for $\query$ and $\db$. 
\end{Definition}

\cite{ps2021} presented a sufficient condition for attribute safety that we utilize in this work. We use $\safeatt(\query)$ to denote the set of attributes that are determined to be safe using the rules of \cite{ps2021}. Any set of safe attributes is a candidate for capturing a provenance sketch. As we discussed in \Cref{sec:introduction}, our goal is to select an optimal (or near optimal) set of attributes to build a sketch on. An additional degree of freedom is the choice of what ranges to use. In this work, we assume the ranges $\ranges_{\att}$ for an attribute $\att$ are provided as input to our technique. For example, a sensible choice is to use the bounds of equi-depth histograms that most databases maintain as statistics.

\subsection{Sketch Selectivity}
\label{sec:sketch-size}

We will use the selectivity of a sketch, i.e., the  amount of data covered by the sketch, as a surrogate for the effectiveness of the sketch to reduce query execution cost.
We measure the \emph{selectivity} \sel of a sketch which is computed for a database $D$, query $\query$,  attribute $\att$, and set of ranges $\ranges$ as shown below. Here \aasize denotes the actual size $\relInst{\provSketch}$ of a sketch \provSketch build for \query on \att over \db and the range partition $\parti_{\ranges,a}(R)$.
$$
\sel(\query,\db,\rel,\att,\ranges) = \frac{\aasize}{\card{\rel}}
$$
%
In the following, we will drop arguments \sizename and similar constructs if they are clear from the context.



\subsubsection{Relative size error}\label{section:rse}

Given an \emph{estimator} \esel of \sel for a provenance sketch, the quality of the estimation is evaluated using the \emph{relative size error} \rse which is computed as shown. 
$$
\rse(\esel) = \frac{\card{\esel - \sel}}{\sel}
$$

 \subsection{Problem Definition}
 \label{sec:problem-definition}

We propose techniques for computing estimates \aesize for candidate provenance sketches for a query $\query$. The purpose of this is two-fold: (i) the size estimate $\esel(\query,\db,\rel,\att,\ranges)$ for a sketch on attribute $\att$ will be used to decide whether it will be worthwhile to build a sketch on $\att$, e.g., a sketch that is estimated to contain 90\% of the data of tables accessed by a query is not effective for improving query performance; (ii) we can rank all viable candidate sketches based on their size estimates to select what sketch to create (the one with the lowest estimated size or top-$k$ sketch candidates to increase the probability that an optimal sketch is selected). For (i) we need to achieve low enough RSE to determine whether a sketch will improve performance while for (ii) we care about the generated ranking, i.e., the RSE has to be low enough to determine the top-$k$ sketches. We will investigate several strategies to select a set of viable attributes for candidate sketches, e.g., attributes that are in some form relevant to a query or all attributes of a table accessed by a query.

\section{Overview \& Problem Definition}
\label{sec:overview}

\begin{figure}
    \centering
    \includegraphics[width=3in]{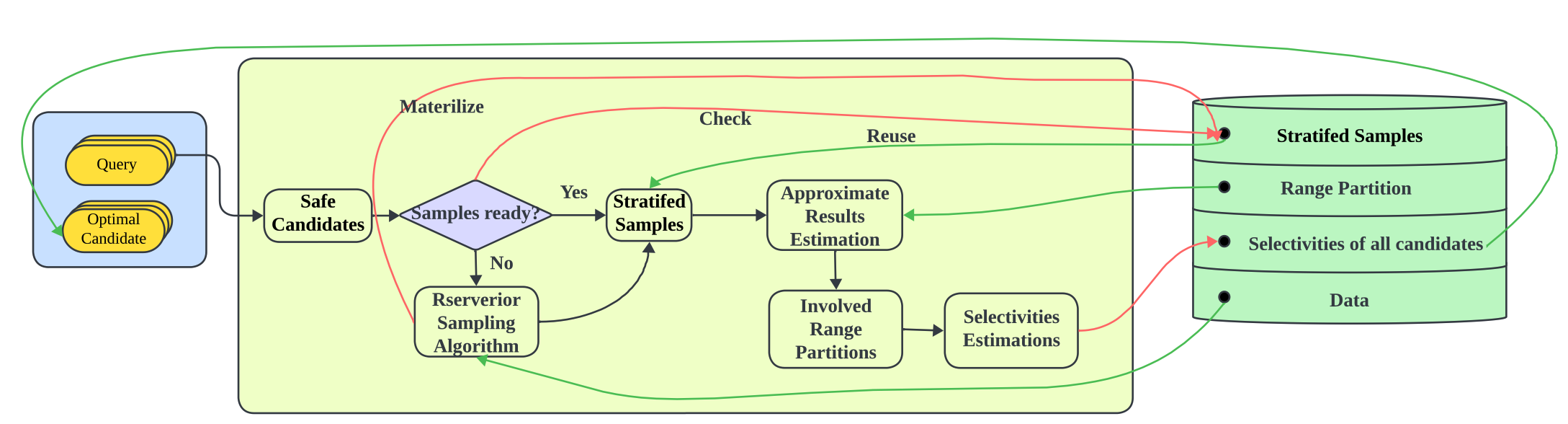}
    \caption{Workflow of cost model}
    \BG{It may be good to show also what state our method saves and also saying optimal candidate is not correct. Plus it is not clear what optimal candidate is. It is not clear from the picture that we search over multiple candidate sketches.}
    \label{fig:work-flow}
\end{figure}

As mentioned in the introduction, the choice of which attribute(s) to build a sketch on can significantly impact the selectivity of the sketch and, thus, the performance of queries using the sketch. In this work, we build a framework that given a query \query and the current state of the database \db automatically selects an attribute from a set of candidate attributes \candA. \Cref{fig:work-flow} show an overview of our framework. Our framework keeps track of existing sketches in an index that allows retrieving a sketch that could be used for a query. \BG{Expand on how this works} Given an input query \query, we first determine whether any existing sketch can be used to answer the query. If that is the case, then we instrument the query to filter data based on the sketch and execute it. For that we use existing techniques from \cite{ps2021} which proposed techniques for determining whether a sketch created for a query $\query_1$ can be used to answer a query $\query_2$. If none of the existing sketches can be used, we employ a \emph{candidate selection strategy} to select for a relation $\rel$ accessed by the query a set of candidate attributes $\candA \subseteq \schemaOf{\rel} \cap \safeatt(\query)$ on which we may want to build a sketch on. Recall that $\safeatt(\query)$ denotes the set of attributes that are guaranteed to yield sketches that produce correct results for $\query$. For queries that access multiple relations we may create several sketches -- each for a different relation.\footnote{A sketch may also be build on a partition over multiple attributes. However, for ease of exposition we single attribute candidates.\BG{Can we back this choice up in experiments?}} We devise several strategies for selecting candidate attributes, e.g., by analyzing which attributes are ``relevant'' for the query. To then select an attribute from the set of candidate attributes we develop a sample-based technique rooted in approximate query processing (\abbrAQP) that estimates the selectivity of a sketch build on a candidate attribute. We utilize this method to select the attribute with the lowest estimated selectivity to build a sketch on (or the top-k attributes to increase robustness against estimation errors). We then build a sketch on the selected candidate attribute(s), add this sketch(es) to the sketch index, and instrument the query to filter data based on the generated sketch.

\parttitle{Candidate Selection Strategies}
As estimating the selectivity of a sketch for a candidate attribute \att is associated with a non-trivial cost, it is sensible to filter out candidate attributes that are unlikely to yield effective sketches. Consider a table \rel with attributes $\schemaOf{\rel} = (\att_1, \ldots, \att_n)$. As mentioned above not all attributes will result in safe sketches, i.e., sketches which guarantee that evaluating the queries over the data described by the sketch will return the same result as evaluating the query over the full database. Thus, any set of candidate attributes $\candA$ has to be a subset of $\safeatt(\query)$.
For each attribute $\att \in \candA$ we assume that the ranges $\ranges_A$ on which we would partition $\att$ on are given as input to our method. For instance, such ranges can be determined based on the bucket boundaries of equi-depth histograms that most DBMS maintain as statistics for query optimization.
While setting $\candA = \safeatt(\query)$ ensures that we do not miss any viable candidate, we consider several strategies for further filtering the candidate set. Specifically, we will analyze the query statically to determine which attributes are relevant for the query, e.g., attributes used in group-by, selection or join conditions, and as input to aggregation functions. The rationale for this choice is that such attributes have a higher chance to be predictive of a row belonging to the provenance of a sketch and, thus, the fragments of the partition for such an attribute better separate provenance from non-provenance resulting in smaller sketches. In our experimental evaluation we compare different strategies including one where $\candA = \safeatt(\query)$ and several strategies that focus on specific types of attributes accessed by a query, e.g., group-by attributes. We will discuss these strategies in more detail in \abbrAQP  \Cref{sec:Candidate attributes and strategies}.

\parttitle{Estimating Sketch Size}
Given a set of candidate attributes \candA, we then want to select the attribute that results in the most selective sketch. A naive approach would be to generate sketches for each candidate $\att \in \candA$ and then select the most selective sketch. However, generating sketches is associated with significant overhead. To reduce this cost, we instead estimate the size of a sketch over a sample of the data. This is akin to approximate query processing (\abbrAQP), but instead of estimating the result of a query we want to estimate, based on samples, the number of fragments that contain provenance which determines the size of the sketch if we take the size of individual fragments into account which can be computed once upfront. We focus on specific types of queries involving aggregation for which \abbrPBDS is very effective. We will discuss the supported classes of queries in \Cref{sec:supp-query-class}. Specifically, we want to estimate the expected number of fragments that will belong to a sketch build on a candidate attribute \att. Recall that a sketch contains all fragments from a relation according to a range partition based on a set of ranges $\ranges$ that contain at least one row of provenance. As an example, consider a simple group-by aggregation query with a \lstinline!HAVING! clause. The provenance for such a query will contain tuples from the input that belong to groups that fulfill the \lstinline!HAVING! condition. The input tuples from each of these groups will possibly contain tuples from multiple fragments of the range partition. To estimate the size of the provenance sketch according to $\ranges$ we have to predict the count of \textit{distinct} fragments that belong to groups that fulfill the \lstinline!HAVING! condition. This requires estimation of the aggregation results for each group to estimate which groups will pass the \lstinline!HAVING! clause. For each such group, the fragments that contain tuples from this group will belong to the sketch.

We now introduce a concrete example that will nonetheless be sufficient for illustrating the challenges of estimating sketch sizes and explain how they are overcome by our approach. Consider the query shown below and suppose we want the estimate the selectivity of a sketch build on attribute \texttt{b} and the range partition $\ranges_b$ we are using contains three ranges $r_1$, $r_2$, and $r_3$.

\begin{lstlisting}
SELECT sum(a) AS sa, b, c
FROM R GROUP BY b, c
HAVING sum(a) > 10
\end{lstlisting}

In this example, the data of each group $(b_1,c_1)$ belongs to a single fragment. Each fragment according to $\ranges_b$ may contain one or more groups.\footnote{In general this is not the case, groups may be spread across multiple fragments for other choices of partition attributes, e.g., if $R$ has another attribute $d$ on which we partition, then the data of a group $(b_1,c_1)$ may be spread across multiple groups.} Using a random sample of $R$, off-the-shelf approximate query processing techniques enable us to compute an estimated value $\tilde{sa}$ of \texttt{sa} for
each group paired with a confidence interval providing the guarantee that with probability at least $\delta$, e.g., 95\%, the actual result will be within $\tilde{sa} \pm \epsilon$.  \BG{Explain that based on the techniques used to calculate bounds, we can then estimate the probability that the value is above the threshold for the HAVING clause} We can then estimate the probability that this group will be in the result of the query and, thus, the fragment it belongs to will be part of the provenance sketch. Let $X_g$ be a Boolean random variable indicting that group $g$ fulfills the \lstinline!HAVING! clause (and, thus, is in the query result) and let us use $p_g$ to denote its probability. To estimate the probability that a fragment containing groups $g_1$ to $g_n$ is in the sketch, we have to estimate the probability of the event $X_{g_1} \vee \ldots \lor X_{g_n}$. As the results for different groups are independent, the probability can be computed as  $1 - ((1 - p_{g_1}) \cdot \ldots \cdot (1 - p_{g_n}))$. However, this approach is a bound to lead to a coarse over-approximation of the probability of a fragment to belong to the sketch due to the fact that the probabilities $p_{g_i}$ are estimates. For instance, let us assume that for each $p_{g_i}$ is normal distributed (based on ways how bounds are calculated). And the central limit theorem  gives a probability of $X_g$ \cite{haas1997large}. The variance and mean statistic of $X_g$ can be generated. Based on the central limit theorem, we can reversely calculate the possibility of that whether the group $g$ fulfill \lstinline!HAVING! clause. The approximating result based on the sample may overestimate or underestimate. The confidence interval are used to avoid this problem. The confidence interval guarantees the expectation have 95\% which is a common confidence level lays in the acceptable intervals.




Let us assume that each fragment contains $1000$ groups and each partition will includes few groups, and we use the group by attributes to generate the provenance sketch. The naive sampling way is direct sampling, sampling on the entire dataset. However, there are some drawbacks in our cases even though it's easy to implement. Based on the direct samples, we can generate the approximating results with 95\% confidence level for each group. However, we still have 5\% possibility to overestimate or underestimate aggregation results. In our case, one group belongs to one partition. As a result, the overestimating result will result in that the partition containing the group be used to calculate the size of partition, which will increase the inaccuracy. What's more, considering the situation that a aggregation query with a huge number of groupby groups, if we directly sample on the entire table with a relatively small sample rate, many groups will be ignored and be represented by no tuple, which result in the inaccuracy of approximating results. To overcome this problems, we first generate the stratified samples \cite{chaudhuri2007optimized} based on the group by attributes if exists, where we implement sampling technique for each groupby groups, making sure each group is represented. Then we estimated the approximated query result according to the stratified samples we generate.

\BGI{******************************************************************************** Previous version below, keep until all relevant information is merged into above}

As already explained in the introduction, which attribute we choose for building a provenance sketch for a query can significantly affect the selectivity of the sketch. The significant factor in determining the optimal candidate is the selectivity associated with the candidate attribute. Selectivity, defined as the proportion of tuples in the provenance sketch relative to the overall size of the dataset, is important. The higher selectivity is associated with higher cost for particular candidate attribute, because more tuples from the input are captured and utilized within the provenance sketch. Selectivity is crucial and direct for us to understand the efficiency for a provenance sketch candidate. The lower selectivity results in the lower cost of the provenance sketch. Our work is choosing the optimal attribute candidate for the provenance sketch of a query, which has the most minimal selectivity in most cases.
\BGI{The following is not clear. Explain why we do stratified sampling and why we stratify on the groupby}

Consider an simple example, we first generate stratified samples ensuring each group represented to increase the accuracy of the estimation. After we generate the stratified samples we calculate the estimation result $SUM\_A$ for each group. Then we apply the selection condition to the approximating results. Supposed we have $(SUM\_A_1,B_1,C_1),((SUM\_A_2,B_2,C_2)$ satisfying the selection. The next step is to calculate the selectivity of each ps candidate. Supposing the B is the candidate, we have the range partition of the B. We check the $B_1,B_2$ belongs to which B range. Then we can calculate the selectivity. Supposing the A is the candidate, we have the range partition of the A. We check the $B_1,B_2$ belongs to which A range by merging the $B_1,B_2$ and R. Then we can calculate the selectivtiy.

To realize this, we first need to generate approximating results to help us estimate the selectivity. The sampling techniques are usually utilized in the approximating processing. The naive sampling way is direct sampling, sampling on the entire dataset. However, there are some drawbacks in our cases even though it's easy to implement. Considering we have 100 partitions and each partition will includes few groups, and we use the group by attributes to generate the provenance sketch. The naive sampling way is direct sampling, sampling on the entire dataset. However, there are some drawbacks in our cases even though it's easy to implement. Based on the direct samples, we can generate the approximating results with 95\% confidence level for each group. However, we still have 5\% possibility to overestimate or underestimate aggregation results. In our case, one group belongs to one partition. As a result, the overestimating result will result in that the partition containing the group be used to calculate the size of partition, which will increase the inaccuracy. What's more, considering the situation that a aggregation query with a huge number of groupby groups, if we directly sample on the entire table with a relatively small sample rate, many groups will be ignored and be represented by no tuple, which result in the inaccuracy of approximating results. To overcome this problems, we first generate the stratified samples \cite{chaudhuri2007optimized} based on the group by attributes if exists, where we implement sampling technique for each groupby groups, making sure each group is represented. Then we estimated the approximated query result according to the stratified samples we generate.

The approximated query result can be used to conduct size estimated. After size estimated, the attribute with minimal\BG{estimated} selectivity can be chosen. 
Given a query, we take the safety check from \cite{ps2021} and apply it here to filter the potential attribute candidates candidates. The range partition for each provenance sketch candidate are created, as introduced in \cite{ps2021}. We divide the entire table into  the specific number of partitions based on the candidate attribute range

\BG{Say briefly how} For each query input, we check if dataset stratified samples, corresponding to that groupby, already exist. If not, we use stratified sampling technique to obtain the necessary samples, stratified based on the group by attributes if exist, making our estimation more accurate as we mentioned before. If the group by attributes don't exist, we implement the sampling technique on the whole table, for example based on row\_id, because we don't need to stratified sample because no group exist. Then we use a bootstrap \cite{tibshirani1993introduction} method to calculate the statistics for the samples, a process of several times resampling from the original sample and merging to get average statistics of original sample to ensure a more accurate the confidence interval. \BG{Need to explain what samples we need and why.}\BGI{This says what samples, but not why} Then the samples statistics are stored for future reuse. If samples statistics are already stored, we directly utilize them. Next, our result estimation algorithm operates on the sampling to approximate the query result. We scale statistics based on the sample size to generate the estimated results. Through our cost model rules\BG{That step is not clear}, \BG{What selection condition?} determine which partitions will be contained in the provenance sketch. Then we estimate the selectivity of the provenance sketch for a given attribute candidate. Once the selectivity is estimated for all candidates, we can select the attribute estimated to yield the sketch for the query with the smallest selectivity. \Cref{sec:cost-model} will give a brief introduction each part of our cost model and define the critical value used in our cost model. The \Cref{section:samplings} will introduce the stratified sampling technique, capturing and use of the stratified samples. The \Cref{sec:selectivity-estimation} will introduce how we estimation the provenance sketch size and calculate the selectivity.  The \Cref{sec:Candidate attributes and strategies} will introduce the strategies we used in the cost model in detail, such as RAND-GB, COST-OPT-GB. The \Cref{sec:Implementation} will introduce how we implement the algorithm we discussed before. And The \Cref{sec:experiments} will introduce how we run the experiments to prove the efficiency and effectiveness of our cost model.

\section{Cost model}\label{sec:cost-model}

\Cref{overview} provides a comprehensive outline of our cost model's workflow. In this section, we introduce our size estimation algorithm for provenance sketches that given a query $\query$, database $\db$, range partitions $\parti_{\ranges,a}(R)$ which can be generated as we introduced in \cite{ps2021}, sample rate $\samplerate$ and set of attributes $\att$ from a relation $\rel$ accessed by the query, estimates the size of a provenance sketch for $\query$ build on $\rel.\att$, which is the total size of each range partitions containing the qualified results 

\BG{We need to explain our assumptions about ranges used for the sketch (are we assuming that ranges are given? How does the estimation depend on the selected ranges.)} model takes a query and datasets as its input and yields the optimal candidate for capturing the provenance sketch.
\BGI{Motivate why these query templates. Why is this is a useful selection? Can we justify this?}

\subsection{Supported Query Classes}
\label{sec:supp-query-class}

Our cost model supports some query class which is shown below. These query templates are common and usually used in the practices and are they are basic type which can be extended further. \QAGH which is aggregation group-by having query, \QAJGH which is aggregation-join-group-by having query and \QAAGH which is aggregation-aggregation-group-by having query templates and \QAAJGH which is aggregation-aggregation-join-groupby-having query templates. The clauses in [] are optional.\\
\noindent\begin{minipage}{0.8\linewidth}
\captionsetup{singlelinecheck = false, justification=justified}
\lstset{upquote=true,frame=single, escapechar=@, title = \bf{Q-AGH: aggregation-groupby-having template }}
\begin{lstlisting}
SELECT @\textcolor{red}{$A_{GB}$,f($A_{agg}$)}@ AS result
FROM R
[WHERE @\textcolor{red}{$A_{GB}$}@]
GROUP BY @\textcolor{red}{$A_{GB}$}@
[HAVING result > @\textcolor{red}{\texttt{\$1}}@]
\end{lstlisting}
\end{minipage}\\
\noindent\begin{minipage}{0.8\linewidth}
\captionsetup{singlelinecheck = false, justification=justified}
\lstset{upquote=true,frame=single, escapechar=@, title = \bf{Q-AJGH: aggregation-join-groupby-having template }}
\begin{lstlisting}
SELECT @\textcolor{red}{$A_{GB}$,f($A_{agg}$)}@ AS result
FROM R,S,..
[WHERE @\textcolor{red}{$A_{GB}$}@]
GROUP BY @\textcolor{red}{$A_{GB}$}@
[HAVING result > @\textcolor{red}{\texttt{\$1}}@]
\end{lstlisting}
\end{minipage}\\
\noindent\begin{minipage}{0.8\linewidth}
\captionsetup{singlelinecheck = false, justification=justified}
\lstset{upquote=true,frame=single, escapechar=@, title = \bf{Q-AAGH: aggregation-aggregation-groupby-having template }}
\begin{lstlisting}
SELECT @\textcolor{red}{$A_{GB2}$,f($A_{agg}$)}@ AS result2
FROM (SELECT @\textcolor{red}{$A_{GB1}$,f($A_{agg}$)}@ AS result1
	 FROM R
	 [WHERE @\textcolor{red}{$A_{GB1}$}@]
	 GROUP BY @\textcolor{red}{$A_{GB1}$}@
	 [HAVING result1 > @\textcolor{red}{\texttt{\$1}}@])
GROUP BY @\textcolor{red}{$A_{GB2}$}@
[HAVING result2 > @\textcolor{red}{\texttt{\$2}}@]
\end{lstlisting}
\end{minipage}\\
\noindent\begin{minipage}{0.8\linewidth}
\captionsetup{singlelinecheck = false, justification=justified}
\lstset{upquote=true,frame=single, escapechar=@, title = \bf{Q-AAJGH: aggregation-aggregation-join-groupby-having template }}
\begin{lstlisting}
SELECT @\textcolor{red}{$A_{GB2}$,f($A_{agg}$)}@ AS result2
FROM (SELECT @\textcolor{red}{$A_{GB1}$,f($A_{agg}$)}@ AS result1
	 FROM R,S,..
	 [WHERE @\textcolor{red}{$A_{GB1}$}@]
	 GROUP BY @\textcolor{red}{$A_{GB1}$}@
	 [HAVING result1 > @\textcolor{red}{\texttt{\$1}}@])
GROUP BY @\textcolor{red}{$A_{GB2}$}@
[HAVING result2 > @\textcolor{red}{\texttt{\$2}}@]
\end{lstlisting}
\end{minipage}\\

Our size estimation algorithm consists of the following steps: a safety check to generate attributes candidates for provenance sketch where we will choose the best attribute among according to their size of provenance sketch estimated, sampling used for result estimation which captures the sample we need to generate the approximating results, result estimation applying approximating query processing technique to generate the qualified results to help us calculate the selectivity, and selectivity estimation to compute the selectivity for each candidate, helping us to determine the best attribute candidate based on the selectivity we  estimated.\BG{You need to explain briefly what the purpose of these steps is. Otherwise, this does not convey any information.} In this section, we will introduce each part of the cost model. It is designed to select the optimal candidate for capturing the provenance sketch for a particular query. The cost model uses size estimation and stratified sampling techniques. Before explaining the individual steps of our size estimation algorithm, we first formally define the selectivity of a provenance sketch and an error metric size estimates.

\subsection{Estimate size} \label{Estimate_size}
Because the selectivity is significant for determining the optimal attribute for provenance sketch. After we define the selectivity given an attribute, estimate the selectivity is the core task of our cost model. We will using sampling technique and wander-join algorithm to estimate the approximating query result $\aqr$. Then the $\aqr$, $\pred{\query}$ and $\parti_{\ranges,a}(R)$ will be used to generate the fragments which we estimated contain the provenance. The The size of fragments which contain the provenance is the critical factor for calculating the selectivity. $\satranges$ denotes the ranges which we estimated contain the provenance. We will introduce in detail $\satranges$ in \cref{sec:Size_estimation}. As a result, $\aesize = \sum_{{r\in \satranges}} \# \rel_{r}$.

\section{Sampling}
\label{section:samplings}
This section introduces the the sampling techniques we use to capture the samples for next step. The input of this section is the datasets and, and the output is stratified samples bootstrapped statistic for each group by.
\BG{Somewhere in the beginning here, you should introduce clearly, e.g., SQL template or state what operators are allowed, the class of queries that is supported / dealt with here} 
Our method is to sample based on the group-by attributes and subsequently do approximate query result of the query\BG{How is this related to provenance sketch size?} on this sample. Once we have the approximate query result, we can join them with range-partitions $\ranges$ to check which fragments contain provenance. Then we can calculate the $\esel(\query,\db,\att,\parti_{\ranges,a}(R))$. We employ the Stratified Reservoir Random Sampling  algorithm for this purpose. One of the primary benefits of the sampling technique is its ability to estimate the approximate result of each group-by with minimal runtime overhead 
\BG{minimal compared to what}. The accuracy of the estimation of this method largely relies on the quality of the samples chosen.

\subsection{Stratified Reservoir Random Sampling Algorithm}
\label{section:Stratified samplings}
\BG{At this point it is not clear how the sample will be used to estimate the selectivity, you need to give an intuition.}
For  queries which contain group-by including\QAGH, \QAJGH, \QAAGH and \QAAJGH, in order to estimate the selectivity, our approach involves capturing stratified samples anchored to the attributes used for group by. We obtain a sample size equivalent to a parameter sample rate $\samplerate$ such as 5\% of the dataset to help us in estimating its statistics. Our sampling on the group attributes is to ensure each group is represented\BG{how do we ensure that for large number of distinct groups?}, increasing the accuracy of our estimations comparing doing reservoir random sampling algorithm on the whole table. It's a trade off: investing more time in capturing samples can increase accuracy, but at the potential cost of increased overhead. If the table large number of distinct groups for example larger than $\samplerate$ times the  total size, we can directly run reservoir random sampling algorithm on the whole table because the if the distinct groups number larger than the sample size, some distinct groups will not be represented. Similarly, while a larger sample size can provide a more precise estimation it comes at the cost of increased estimation runtime. We now define stratified samples.

\begin{Definition}[Stratified samples]
  Consider a relation $\rel$, $\att $, set of $\att$ denoted as $\attset_{gb} \in \relSchema$. Let $\group{\attset_{gb}}$ denote the set of all distinct groups values on $\attset_{gb}$ .Each group can be denoted by a unique group identifier \GID. Let $g_{\GID}$ denote each distinct groups value, such that $\group{\attset_{gb}} = \{g_1, \ldots, g_n\}$. Let $\samples{\group{\attset_{gb}}}{\samplerate} = \{s_1, \ldots, s_n\}$ be a set of samples generated by \emph{reservoir random sampling} based on group bys $\group{\attset_{gb}}$ with the sample rate $\samplerate$, where each tuple has $\samplerate$ probability to be chosen, such that $\bigcup_{{\GID}=0}^{n} s_{\GID} = \samples{\group{\attset_{gb}}}{\samplerate}$ and $s_{\GID} \subseteq \{ t \mid t \in R \wedge t.a = g_{\GID} \}$
\end{Definition}

Consider the example for generating generate stratified samples shown in \Cref{samples_crimes} which uses the same data as  shown in \Cref{fig:CRIMES}. These samples are captured based on the group by attributes district and zip\_codes, resulting in groups g1, g2, and g3. Within each group, we deploy the Reservoir Random Sampling Algorithm\BG{Why?} to selectively generate samples. The reservoir random sampling algorithm can ensure each tuple has the same probability to be chosen within O(n) time complexity where n is the total size of table. \BG{What is $n$ here?} The algorithm may chose, e.g., the rows highlighted in red in \Cref{sample}.

If samples of $\query_1$ based on group-by have been captured and $\query_2$ share the same group-by attributes and distinct values of group-by in $\query_2$ is a subset of that in$\query_1$, the samples of $\query_1$ can be reused. Hence, our approach caches generated samples, anticipating reuse for queries with the same group-by attributes. 
For the query without group-by, we directly run reservoir random sampling without stratification.

\subsection{Bootstrap sampling}
\begin{figure}
    \centering
    \includegraphics[width=3in]{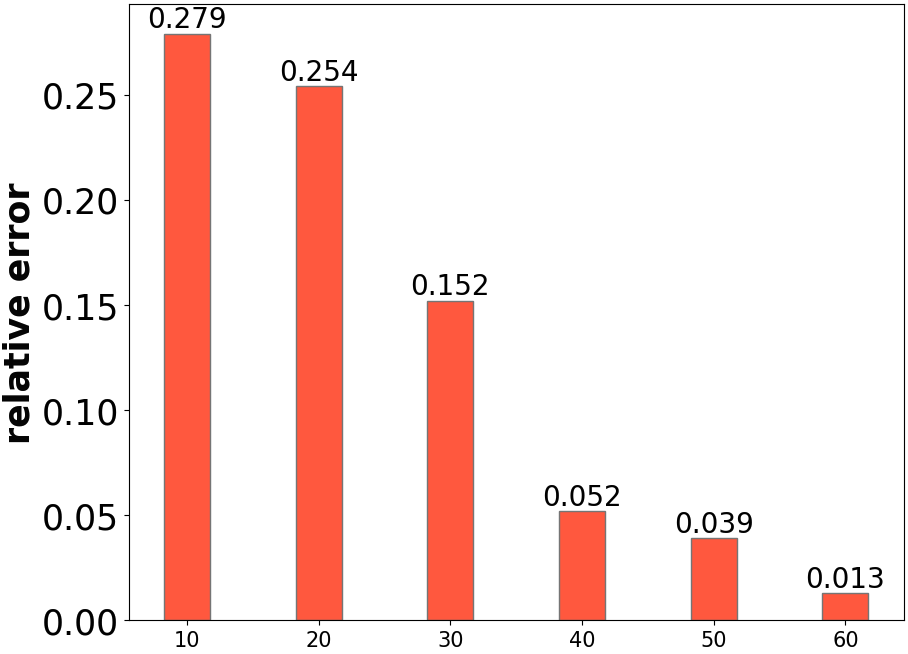}
    \caption{Relative sketch data size error varying times of resamples over TPCH}
    \label{relative_error_different_resamples}
\end{figure}

After we generate the stratified samples we discussed in the \ref{section:Stratified samplings}. In order to increase the accuracy of estimation and to provide confidence intervals, we use a bootstrapping procedure \cite{tibshirani1993introduction} which only adds minimal computational overhead. Bootstrap uses random sampling with replacement for resampling. Bootstrapping assigns measures of accuracy to sample estimates. In our practice, the samples $s_{\GID}$ based on the $GID$ is generated by the Stratified Reservoir Random Sampling Algorithm. For each $GID$, From $s_{\GID}$, resamples are generated by drawing with replacement randomly for the same sample size. According to the \cite{tibshirani1993introduction}, we repeat the process for a specific times, which have a relative good accuracy with low overhead. The more times we resample, the more accuracy estimation we will get, while more overhead we have. We plot the relative error for different resample times in Fig. \ref{relative_error_different_resamples}. From the plot, we could see resample 50 times can have relative low error. For instance, suppose we generated $s_1$ with 30 tuple size. Then we randomly generate 30 tuples from $s_1$ as a resample $rs_1$ and repeat 50 times. $rs_1, rs_2, ... , rs_{50}$ are used to estimate the statistic of the population. As a result, the statistic for $s_i$ is denoted as $ \overline{s_i} = \frac{\sum^{50}_{1} \overline{rs_i}}{50}$

\begin{figure}[h]

    \begin{subfigure}{0.9\linewidth}
        \captionsetup{singlelinecheck = false, justification=justified}
        \lstset{upquote=true,frame=single, title = \bf{\qhighcrimes}}
        \begin{lstlisting}
        SELECT sum(records) AS
       	num_crimes,pid,month,year
        FROM crimes
        GROUP BY pid,month,year
        HAVING num_crimes >= 100;
        \end{lstlisting}
        \caption{Running example query}
        \label{fig-example}
    \end{subfigure}'

     \begin{subfigure}{0.9\linewidth}

        \begin{tabular}{c|c|c|c|c|}
        \hhline{~|-|-|-|-|}
          &\cthead pid & \cthead month & \cthead year & \cthead records\\
       	$g'_0$	&	 \color{LRed}{3}               & \color{LRed}{1}       & \color{LRed}{2010}  & \color{LRed}{88}    \\
		$g'_1$   &  \color{LRed}{\textbf{4}}       & \color{LRed}{\textbf{1}}       & \color{LRed}{\textbf{2013}}   & \color{LRed}{\textbf{101}} \\
		$g'_2$ &  \color{LRed}{\textbf{8}}       & \color{LRed}{\textbf{6}}       & \color{LRed}{\textbf{2015}}   & \color{LRed}{\textbf{86}}     \\
        $g'_3$&  \color{LRed}{\textbf{2}}       & \color{LRed}{\textbf{7}}       & \color{LRed}{\textbf{2016}}   & \color{LRed}{\textbf{157}}    \\
        $g'_4$&	\color{LRed}{7}       & \color{LRed}{2}      & \color{LRed}{2022}   &\color{LRed}{83} \\
       	$g'_5$&	\color{LRed}{7}       & \color{LRed}{9}     & \color{LRed}{2023}   &\color{LRed}{58}   \\
       & ...    & ...             & ...          & ...                                    \\
       \hhline{~|-|-|-|-|}
        \end{tabular}
        \caption{samples}
        \label{sample}

    \end{subfigure}

    \caption{Example for Stratified Sampling on Group-by Attributes}
  \end{figure}

   \begin{figure}[h]
\begin{subfigure}{0.8\linewidth}
         \begin{tabular}{c|c|c|c|c|}
        \hhline{~|-|-|-|-|}
          &\cthead pid & \cthead month & \cthead year & \cthead records\\
       	$g'_0$	&	 \color{LRed}{3}               & \color{LRed}{1}       & \color{LRed}{2010}  & \color{LRed}{88}    \\
		$g'_1$   &  \color{LRed}{4}       & \color{LRed}{1}       & \color{LRed}{2013}   & \color{LRed}{202} \\
		$g'_2$ &  \color{LRed}{8}       & \color{LRed}{6}       & \color{LRed}{2015}   & \color{LRed}{172}     \\
        $g'_3$&  \color{LRed}{2}       & \color{LRed}{7}       & \color{LRed}{2016}   & \color{LRed}{157}    \\
        $g'_4$&	\color{LRed}{7}       & \color{LRed}{2}      & \color{LRed}{2022}   &\color{LRed}{83} \\
       	$g'_5$&	\color{LRed}{7}       & \color{LRed}{9}     & \color{LRed}{2023}   &\color{LRed}{58}   \\
       & ...    & ...             & ...          & ...                                    \\
        \hhline{~|-|-|-|-|}
        \end{tabular}
        \caption{group estimation}
        \label{group_estimation}
    \end{subfigure}

    \begin{subfigure}{0.8\linewidth}
        \begin{tabular}{c|c|c|c|c|}
        \hhline{~|-|-|-|-|}
          &\cthead pid & \cthead month & \cthead year & \cthead records\\
		$g'_1$   &  \color{LRed}{4}       & \color{LRed}{1}       & \color{LRed}{2013}   & \color{LRed}{202} \\
		$g'_2$ &  \color{LRed}{8}       & \color{LRed}{6}       & \color{LRed}{2015}   & \color{LRed}{172}     \\
        $g'_3$&  \color{LRed}{2}       & \color{LRed}{7}       & \color{LRed}{2016}   & \color{LRed}{157}    \\
         \hhline{~|-|-|-|-|}
        \end{tabular}
        \caption{satisfied groups}
        \label{satisfied_groups}
    \end{subfigure}

    \begin{subfigure}{0.8\linewidth}
        \begin{tabular}{c|c|c|c|c|}
        \hhline{~|-|-|-|-|}
          &\cthead pid & \cthead month & \cthead year & \cthead records\\
		$g'_1$   &  \color{LRed}{4} \cellcolor{shadegreen} & \color{LRed}{1} \cellcolor{shadegreen} & \color{LRed}{2013} \cellcolor{shadegreen} & \color{LRed}{202} \cellcolor{shadegreen} \\
		$g'_2$ &  \color{LRed}{8} \cellcolor{shadered}   & \color{LRed}{6}\cellcolor{shadered}  & \color{LRed}{2015}\cellcolor{shadered}& \color{LRed}{172}\cellcolor{shadered}  \\
        $g'_3$&  \color{LRed}{2} \cellcolor{shadeblue}   & \color{LRed}{7}\cellcolor{shadeblue} & \color{LRed}{2016}\cellcolor{shadeblue} & \color{LRed}{157}\cellcolor{shadeblue} \\
         \hhline{~|-|-|-|-|}
        \end{tabular}
        \caption{satisfied groups in pid provenance sketch}
        \label{satisfied_groups_pid}
    \end{subfigure}

	\caption{Selectivity Estimation}
    \end{figure}

\section{Selectivity Estimation}\label{sec:selectivity-estimation}
This section introduces how we compute the provenance sketch size of each candidate. The input of stratified samples bootstrapped statistic for each group by, and the output before 8.1 is $\satgroup{\attset_{gb}}$ which is the groupby sets satisfying the conditions. The input of 8.1 $\satgroup{\attset_{gb}}$ and the output is provenance sketch size.


Once we have generated the appropriate samples for the query \QAGH, \QAJGH \QAAGH and \QAAJGH template, or have determined that stored samples can be reused, our subsequent step involves estimating the query result and confidence intervals. Initially, we have the statistics of each group, focusing on metrics such as the number of tuples in each group and aggregate results like SUM, AVG and COUNT. With these statistics, assuming uniform distribution, we can then apply specific rules to approximate the statistics of each group for the entire dataset. The rules are shown below:
\begin{Definition}[Approximating results rules]
If there is no \join in \query.
As we introduce in \Cref{section:samplings}, $g_{\GID}$ denotes the distinct group-by values and $s_{\GID}$ denotes the stratified samples based on $g_{\GID}$.  $\overline{s_i}$ denotes the statistic in $g_{\GID}$. Let $\#g_{\GID}$ denote the number of tuples in group $g_{\GID}$ in the $\rel$ and $\#s_{\GID}$ denote the number of tuples in sample $s_{\GID}$ in the $\rel$. $\#s_{\GID_{pred}}$ denotes the number of tuples that satisfy the predicate in the sample.. $t$ denotes the tuples. \predicate denotes the WHERE predicate of the aggregate query, where \predicate = 1 or 0 denotes t satisfies or dissatisfies the predicate, respectively. The unbiased estimator for aggregation result of $\rel$ on $g_{\GID}$ denoted as $EST_{\aggregation(g_{\GID})}$ is defined as \cite{acharya2000congressional} \cite{liang2021combining} :
\BG{Where do these formulas come from? What do they guarantee?}
  \begin{align*}
     EST_{\aggregation(g_{\GID})} = \frac{1}{\#s_{\GID}}\sum_{t\in s_{\GID}} \phi(t) \\
  \end{align*}

 where $\phi(t)$ expresses all of the necessary scaling factor.
 \begin{itemize}
\item For \aggregation = SUM, $  \phi(t) =  \predicate \cdot \#g_{\GID}  \cdot \overline{s_i}$
\item For \aggregation = AVG, $ \phi(t) = \predicate \cdot \frac{\#s_{\GID}}{\#s_{\GID_{pred}}} \cdot \overline{s_i}$
\item For \aggregation = COUNT, $ \phi(t) =  \predicate \cdot \#g_{\GID}$

\end{itemize}

The approximated query result is denoted as:
   \begin{align*}
   	 \aqr = \{EST_{\aggregation(g_1)}, \ldots, EST_{\aggregation(g_n)}\}
   \end{align*}
 \end{Definition}

 If there is \join in \query, $\aqr$ can be generated by wander-join algorithm\cite{li2016wander}. For the AVG, COUNT,SUM queries with join and selection predicates, confidence interval formulas have been studied by \cite{haas1997large}.\\
 After we get the approximating query result $\aqr$, we need to generate which distinct group-by satisfy the query predicates $\pred{\query}$. Let $\satgroup{\attset_{gb}}$ denote the distinct group-by values which satisfied the query predicates $\pred{\query}$ in \QAGH, \QAJGH,  \QAAGH and \QAAJGH. Then $\satgroup{\attset_{gb}} = \{ g \mid g \in \group{\attset_{gb}} \wedge g \in \aqr$ satisfying $ \pred{\query} \} $. Then $\satgroup{\attset_{gb}}$ will be used to calculate the selectivity, which we will introduce in the next section. The pseudocode code are shown in \cref{alg:aqr} to shows how we compute the approximate query result.
\begin{algorithm}[t]
  \caption{\textsc{Approximate query result} Function}
  \begin{algorithmic}[1]

    \Procedure{AQR}{$\samples{\group{\attset_{gb}}}{\samplerate},\group{\attset_{gb}},Q$}
      \If{$Q \in \QAJGH \mid Q \in \QAAJGH $}
      \State $\aqr \gets \Call{WANDER-JOIN Algorithm}{Q}$\cite{li2016wander}
      	\If{$\pred{\query} \in Q$}
      	\State $\aqr_{sat} \gets  \aqr$ satisfying $ \pred{\query}$
      	\State $\satgroup{\attset_{gb}} \gets\{ g \mid g \in \group{\attset_{gb}} \wedge g \in \aqr_{sat} \} $
      	\State \Return $\aqr_{sat}, \satgroup{\attset_{gb}}  $
      	\Else
      	\State \Return $\aqr, \group{\attset_{gb}}  $
      	\EndIf
      \EndIf

%
%

      	\If{$Q \in \QAGH \mid Q \in \QAAGH $}
       	\State $EST_{\aggregation(g_{\GID})} = \frac{1}{\#s_{\GID}}\sum_{t\in s_{\GID}} \phi(t) $
        \If{$\aggregation = SUM$}
        \State $EST_{\aggregation(g_{\GID})}  =\frac{\predicate \cdot \#g_{\GID} \cdot \overline{s_i}}{\#s_{\GID}}\sum_{t\in s_{\GID}}$

      	\EndIf
		\If{$\aggregation = AVG$}
        \State $ EST_{\aggregation(g_{\GID})} =\frac{\predicate \cdot \overline{s_i}}{\#s_{\GID_{pred}}}\sum_{t\in s_{\GID}}$

      	\EndIf
      	\If{$\aggregation = COUNT$}
        \State $EST_{\aggregation(g_{\GID})} =\frac{\predicate \cdot \#g_{\GID}}{\#s_{\GID}}\sum_{t\in s_{\GID}} $
      	\EndIf

		\State $\aqr \gets \{EST_{\aggregation(g_1)}, \ldots, EST_{\aggregation(g_n)} \}$
		\If{$\pred{\query} \in Q$}
      	\State $\aqr_{sat} \gets  \aqr$ satisfying $ \pred{\query}$
      	\State $\satgroup{\attset_{gb}} \gets\{ g \mid g \in \group{\attset_{gb}} \wedge g \in \aqr_{sat} \} $
      	\State \Return $\aqr_{sat}, \satgroup{\attset_{gb}}  $
      	\Else
      	\State \Return $\aqr, \group{\attset_{gb}}  $
      	\EndIf

      \EndIf

    \EndProcedure
  \end{algorithmic}
  \label{alg:aqr}
\end{algorithm}
\\
To elucidate, consider the continued example from \Cref{sec:introduction}. We draw the stratified samples from the \Cref{fig:CRIMES}, shown in the \Cref{sample}. From this sample, we extract certain statistics.  For instance, referring to the prior example where samples are detailed in \Cref{sample}, our calculations would be as follows:, $EST_{SUM(g_0)} =  \overline{s_0} * \#g_0 = 88 * 1 = 88$, $EST_{SUM(g_1)} = \overline{s_1} * \#g_1 = 101*2 = 202$, $EST_{SUM(g_2)} = \overline{s_2} * \#g_2 = 86 * 2 = 172$, $EST_{SUM(g_3)} = \overline{s_3} * \#g_3 = 157 * 1 = 157$, $EST_{SUM(g_4)} = \overline{s_4} * \#g_4 = 83 * 1 = 83$, $EST_{SUM(g_5)} = \overline{s_5} * \#g_5 = 58 * 1 = 58$. These estimations are graphically depicted in \Cref{group_estimation}.
Having derived the estimated results for each group, we then apply the selection condition to determine which groups meet the specified conditions in \QAGH, \QAJGH, \QAAGH and \QAAJGH. In our example, the selection condition is \textbf{HAVING num\_crimes >= 100}. Applying this to our group estimations, we discern that groups $g'_1$, $g'_2$ and $g'_3$ meet the conditions shown in \Cref{satisfied_groups} . As a result, $\satgroup{\attset_{gb}} = \{g'_1,g'_2,g'_3\}$ will be used to calculate the selectivity. 

\subsection{Size estimation}\label{sec:Size_estimation}
After we get the the distinct group-by values $\satgroup{\attset_{gb}}$ which satisfied the query predicates $\pred{\query}$, given the range partitioning of the table on $\att$, $\parti_{\ranges,a}$, our next step is estimating selectivity involves identifying the fragments that encompass these groups. To achieve our objective, we need to join the range partition of the relation with the estimated results. This allows us to identify the estimated partitions that contain the provenance of the query. After we estimate the partitions which contain the provenance. We can calculate the size of the provenance sketch. The selectivity is defined in the \Cref{sec:sketch-size}.
\begin{Definition}[Size estimation rules] As we introduce in \Cref{section:samplings} and \Cref{sec:selectivity-estimation}, $\group{\attset_{gb}}$ denotes the distinct group-by values. $\satgroup{\attset_{gb}}$ denote the distinct group-by values which satisfied the query predicates $\pred{\query}$. The distinct value set of $\att$ we captured provenance sketch which satisfies the $\pred{\query}$ is denoted as $\satps{\satgroup{\attset_{gb}}}{\db} = \{ t.a \mid t.\attset_{gb}\in \satgroup{\attset_{gb}} \wedge t \in \db \}$. So the satisfied ranges $\satranges = \{ r \mid t\in \satps{\satgroup{\attset_{gb}}}{\db} \wedge r \in \ranges \wedge t \in  r \} $.  Then \aesize we defined in \cref{section:selectivity_ps} can be generate: $\aesize = \sum_{{r\in \satranges}} \# \rel_{r}$.

%
%

\end{Definition}
 The pseudocode code are shown in \cref{alg:selectivity} to shows how we compute the selectivity.
In the context of our example, as illustrated in \Cref{sec:introduction}, the range partition for PID is denoted by $\parti_{pid}$. Within this, $p_1 = [1,3]$, $p_2 = [4,6]$, $p_3 = [7,9]$. For MONTH, the partition is represented $\parti_{month}$ where $m_1 = [1,4]$, $m_2 = [5,8]$, $m_3 = [9,12]$. For YEAR, the partition is represented $\parti_{year}$ where $y_1 = [2010,2012]$, $y_2 = [2013,2020]$, $y_3 = [2021,2024]$.

By referencing the results displayed in \Cref{satisfied_groups} and the range partitions of CRIMES, we infer the following: For the PID candidate, the range partitions $p_1$,$p_2$ and $p_3$ are involved. Conversely, for the MONTH candidate, $\satranges$ = $m_1$, $m_2$. For the YEAR candidate, $\satranges$ = $y_2$.  Identifying the relevant fragments allows us to precisely estimate the selectivity of the provenance sketch. The selectivity is what we are estimating, and is defined in \Cref{sec:sketch-size}.
Based on the defined selectivity of the provenance sketch, our example indicates that the selectivity of the PID candidate stands at $8/8=100\%$ shown in \Cref{satisfied_groups_pid}. In contrast, the selectivity of the MONTH candidate is $7/8=87.5\%$ and the selectivity of the YEAR candidate is $5/8=62.5\%$. Therefore, the YEAR emerges as the optimal candidate for capturing the provenance sketch. To summarize the process for selectivity estimation:
\begin{itemize}[noitemsep,topsep=0pt,parsep=0pt,partopsep=0pt,leftmargin=*]
\item Begin by capturing a representative sample of the dataset.
\item Proceed to estimate the results for each group by scaling the sample estimation outcome.
\item Apply the selection condition to the estimated results and utilize the range partitions to identify the corresponding involved range partitions.
\item Finally, evaluate the selectivity of every candidate to discern the optimal one for the capturing provenance sketch.
\end{itemize}
\BG{You are not explaining the intuition of why this would work / give a reasonable estimation}
\begin{algorithm}[t]
  \caption{\textsc{size estimation}}
  \begin{algorithmic}[1]

    \Procedure{Size estimation}{$\query,\db,\parti_{\ranges,a}(R), \satgroup{\attset_{gb}}$}
      \State $\satps{\satgroup{\attset_{gb}}}{\db} \gets \{ t.a \mid t.\attset_{gb} \in \satgroup{\attset_{gb}} \wedge t \in \db \}$
      \State $\satranges \gets \{ r \mid t\in \satps{\satgroup{\attset_{gb}}}{\db} \wedge r \in \ranges \wedge t \in  r \}  $
      \State $\aesize = \sum_{{r\in \satranges}} \# \rel_{r}$
    \EndProcedure
    \State \Return $\aesize$
  \end{algorithmic}
   \label{alg:selectivity}
\end{algorithm}

\subsection{Selectivity Expectation}\label{sec:Selectivity_expectation}
In this subsection we show how we apply probability theorems, derived formulas and the estimated size we generate in \cref{sec:Size_estimation} to compute the size expectation. As we introduced in \Cref{sec:AQP}, we first calculate the confidence intervals for aggregations\cite{haas1997large} as follows:
We are given a table of $N$ tuples, where each tuple t is associated with value $v(t)$, as well as an indicator variable $u(t)$ that is 1 if $t$ meets the predicate and 0 otherwise. suppose we have sample n tuples randomly, $t_1, t_2,...,t_n$. For any aggregation function $f, h$ defined on $\{1,2,. . . ,m\}$, introduce the following notation:
 \begin{align*}
&\theta(f) = \frac{1}{m}\sum^m_{i=1} f(t_i), \tag{1} \label{eq:1}\\
&T_n(f) = \frac{1}{n}\sum^n_{i=1} f(t_i), \tag{2}  \label{eq:2}\\
&T_{n,q}(f) = \frac{1}{n-1}\sum^n_{i=1} (f(t_i)-T_n(f))^q, \tag{3}  \label{eq:3}\\
&T_{n,q,r}(f,h) = \frac{1}{n-1}\sum^n_{i=1} (f(t_i)-T_n(f))^q(h(t_i)-T_n(h))^r, \tag{4}  \label{eq:4}\\
\end{align*}
\Cref{eq:1} is the aggregation function in the form of an average. With these definitions, the final answer $\mu$ to the query can be written as $\mu=\theta(v)$. After n tuples have been retrieved, the running aggregate is given by $ \overline Y=T_n(v)$ where $T_n(f)$ is shown in \Cref{eq:2}. For $q\geq 0$, a large-sample confidence interval for $ \overline Y$ follows from the strong law of large numbers for random variables that $T_{n,2}(v)\rightarrow \theta_2(v)$ where $T_{n,q}(f)$ is shown in \Cref{eq:3}. Denote by $S \subseteq R$ the set of tuples that satisfy predicate. For $q,r \geq 0$, the large-sample confidence intervals is $\theta_{q,r}(f,h) = \frac{1}{m}\sum^m_{i=1} (f(t_i)-\theta(f)^q(h(t_i)-\theta(h))^r$. Spectively, we have $T_{n,q,r}(f,h)$ shown in \Cref{eq:4}.
\BG{What are $T_n$, $T_{n,q}$, $T_{n,q,r}$ intuitively, you need to say that}

Haas\cite{haas1997large} also derived the estimators for various aggregation functions, as well as estimators for their variances, as follows:
\begin{align*}
&\texttt{SUM}: \ \widetilde{Y}_n=T_n(uv),\ \widetilde{\sigma}_n^2=T_{n,2}(uv); \tag{5} \label{eq:5}\\
&\texttt{COUNT}: \ \widetilde{Y}_n=T_n(u),\ \widetilde{\sigma}_n^2=T_{n,2}(u); \tag{6} \label{eq:6}\\
&\texttt{AVG}: \ \widetilde{Y}_n=T_n(uv)/T_n(u), \\
&\ \widetilde{\sigma}_n^2= \frac{1}{T_n^2(u)}\left(T_{n,2}(uv)-2R_{n,2}T_{n,1,1}(uv,u)+R_{n,2}^2T_{n,2}(u)\right); \tag{7} \label{eq:7}
\end{align*}
where $R_{n,2}=T_n(uv)/T_n(u)$.\\
Finally, we compute $\widetilde{\sigma}_n^2$ according to \Cref{eq:5,eq:6,eq:7}. Then the half-width of the confidence interval can be computed as (for a confidence level threshold $\alpha$)
\begin{align*}
\epsilon_n = \frac{z_\alpha \widetilde{\sigma}_n}{\sqrt{n}}
\end{align*}
where $z_\alpha$ is the $\frac{\alpha+1}{2}$-quantile of the normal distribution with mean 0 and variance 1.\\

\begin{Definition}[Selectivity Expectation] Assuming tuples in the provenance of two groups do not overlap and all values in the confidence interval fulfill the selection criterion or none of them in the query. Let $p$ denotes the confidence level. $X_i$ is the random variable representing the fact that is $g_{\GID}$ in the provenance. $R_i$ is the random variable representing the fact that $r_i \in \provSketch$. Thus, $X_{\GID} = \{p : true; 1-p :false \}$. As we introduced in \cref{Estimate_size}, $\satranges$ denotes the ranges which we estimated contain the provenance. The groups containing provenance denotes as $ \{ G_i \mid t.a \in \rel_{r_i} \wedge t.\attset_{gb} \in G_i  \wedge r_i \in \satranges \}$. The probability of fragments are covered in the provenance sketch $P(r_i \in \provSketch) $ is :
\end{Definition}
  \begin{align*}
   	 P(r_i \in \provSketch) = P(R_i) =P(X_{g_1} \union X_{g_2}\union \ldots \union X_{g_n}) \hspace{1mm}\mathtext{where}\hspace{1mm}  g_1, g_2, \ldots , g_n \in G_i
  \end{align*}
If $g1,g2,\ldots,gn$ is independent.
  \begin{align*}
   	 P(r_i \in \provSketch) = 1- \overline{P(X_{g_1})} \cdot \overline{P(X_{g_2})}\cdot \ldots \cdot \overline{P(X_{g_n})}  = 1- (1-p)^n
  \end{align*}
If $g1,g2,\ldots,gn$ is dependent. According to the Fréchet inequalities, $max_k\{P(A_k)\} \leq P(\union_{k=1}^n A_k) \leq min(1, \sum_{k=1}^n P(A_k))$, considering the independent cases,
 \begin{align*}
   	 max_n(P(X_{g_i})) \leq P(r_i \in \provSketch)  \leq 1-  \prod_1^n \overline{P(X_{g_i})} = 1- (1-p)^n
  \end{align*}

  Let E(\aesize) denotes the expectation size of provenance sketch over candidate \att:

 \begin{align*}
E(\aesize)& = E(\#R_{r_i} \cdot \sum_{{r_i \in \satranges}} R_i ) \\
& = \#R_{r_i} \cdot (E(R_1)+E(R_2)+ \ldots + E(R_i)) \\
& = \#R_{r_i} \cdot  \sum_{{r_i \in \satranges}} P(r_i \in \provSketch)
\end{align*}

As a result, $E(\aesize) \in$
$$[\#R_{r_i} \cdot  \sum_{{r_i \in \satranges}}  max_n(P(X_{g_i})),  \#R_{r_i} \cdot  \sum_{{r_i \in \satranges}} (1-  \prod_1^n \overline{P(X_{g_i})} = 1- (1-p)^n) ]$$

If considering confidence interval overlap the section criterion, $X_{\GID} = \{p \cdot \lambda : true; 1-p\cdot \lambda :false \}$, p is the portion which fulfill selection the condition. If we use normal distribution, $\lambda =\mathcal{Z}_{p/2}(\frac{X - \aqr}{\sigma})$ where $X$ is the value in selection condition.

\section{Candidate Attribute Selection Strategies}
\label{sec:Candidate attributes and strategies}

\BGI{We need a more comprehensive description of strategies, this just prefilters attributes with low number of distinct values}
For the most cases, the performance of the attribute which has small distinct value number is much worse than the performance of the attribute which has large distinct value number. Considering a specific query, if attribute has the more distinct value comparing the other attributes with fewer distinct value, for the same partitions number, the attribute with more distinct value can be divided more accurate, resulting in fewer partitions which contain provenance. In this situation, the selectivity can be smaller. Thus, for our experiments, we firstly have a pre-filtered of the candidates of attribute whose distinct value number is larger than the number range partitions we divided, \textbf{because if one attribute candidate's distinct value number is lower than the number of range partitions, the provenance sketch may generate the wrong result according to the safety rules\cite{ps2021}}\BG{This needs more explanations:} What's more, in our provenance-based data skipping paper \cite{ps2021}, we introduce the safety rules. We apply safety rules to the  pre-filtered attributes. Then the candidates generated for latter estimation.
Many strategies can be applied to choose the attribute for building a provenance sketch for a query. Different strategies can result in different performance of provenance sketch. Different strategies  include random picking from all attributes candidates after per-filtering (\stratRANall), randomly selecting from query-related attributes (\stratRANrelall), randomly selecting a group-by attribute (\stratRANgb), randomly selecting a primary key attribute (\stratRANpk). randomly selecting one of the aggregation input attribute  (\stratRANagg). These strategies utilizing random choice pick an attribute  uniformly at random from a set of candidate attributes specific to the strategy. For example, for randomly selecting a primary key attribute for a relation with a primary key $\{a,b\}$ we would pick each attribute with 50\% probability. In addition to the strategies that pick random attributes, we also evaluate strategies that use our size estimation techniques to select on attribute from a list of candidate attributes.
The \stratCBOPTrel strategy picks the attribute from the set of all safe attributes that is estimated to yield the smallest provenance sketch. The \stratCBOPTgb strategy applies the same selection criterion, but only considers group-by attributes. Because the samples is captured based on the group by attribute, capturing provenance sketch on group by attribute is consistent with the sampling procedure.

\section{Implementation}\label{sec:Implementation}
We implement the stratified sampling method use server programming interface in the postgresql. The Server Programming Interface (SPI) gives writers of user-defined C functions the ability to run SQL commands inside their functions or procedures. SPI is a set of interface functions to simplify access to the parser, planner, and executor. SPI also does some memory management. SPI can speed up the process of our stratified sampling. We sorting the table based on the group by attribute first. Then we use SPI capture the stratified samples and store them for reusing. For the size estimation. We implement it in our provenance system GProm by postgresql.

\section{Experiments}
\label{sec:experiments}
In our experimental evaluation we focus on two aspects. First, we evaluate the accuracy and runtime of our cost model. Then we compare the different strategies can be applied. At last, we conduct end-to-end experiment to evaluate how we benefit from the cost model.
\BGI{Meed a short introductory paragraph here}


\subsection{Experimental setup}
\subsubsection{Workloads}
\BGI{Need to add information about the number of attributes as this is also very relevant for our approach}
\BGI{add citation or links for all datasets}

\parttitle{Chicago Crime} This dataset records crimes reported in Chicago (\url{https://data.cityofchicago.org/Public-Safety/Crimes-2001-to-present/ijzp-q8t2}). It contains $\sim$6.7M tuples and 9 numeric attributes each corresponding to a single crime. 

\parttitle{TPC-H} The TPC-H is a decision support benchmark. It consists of a suite of business oriented ad-hoc queries and concurrent data modifications. It has numeric 10 attributes and 6.15M tuples

\parttitle{Parking} This dataset records parking reports in New York City. It contains $\sim$31M tuples and 16 numeric attributes.

\parttitle{Stars} We obtained this dataset from SDSS-V. SDSS-V is the first facility providing multi-epoch optical \& IR spectroscopy across the entire sky, as well as offering contiguous integral-field spectroscopic coverage of the Milky Way and Local Volume galaxies. It contains $\sim$5.2M tuples and 7 numeric attributes.

\parttitle{Synthetic queries} \BG{Need to motivate why we used synthetic queries and why we generated templates in this way}
Given that these three templates represent the most frequently occurring queries, once a provenance sketch for a query is generated, it can be stored for future use. If subsequent queries in the workload match a previously captured provenance sketch, this stored sketch can be directly reused. For these reasons,
we generated 1000 queries for which is based on three templates over the \dsCrime, \dsTPCH, \dsParking and \dsStars datasets to simulate a practical workload. The queries below are examples of the templates selection-aggregation-groupby, selection-aggregation-join-groupby and complex query templates shown below. The part of the template we are varying is highlighted in red. We randomly select the attributes to group on (\textcolor{red}{$A_{GB}$}), the attribute to aggregate over (\textcolor{red}{$A_{agg}$}), the aggregation function (\textcolor{red}{$SUM$}),(\textcolor{red}{$AVG$}) and a threshold for the selection condition (\textcolor{red}{\$1}).
We run these queries on our cost model to test the effectiveness and accuracy. These templates are safe for our Provenance Sketch and encompass most of our routine use cases.

\noindent\begin{minipage}{0.99\linewidth}
\captionsetup{singlelinecheck = false, justification=justified}
\lstset{upquote=true,frame=single, escapechar=@, title = \bf{\QAGH : groupby-having template }}
\begin{lstlisting}
SELECT @\textcolor{red}{$A_{GB}$,f($A_{agg}$)}@ AS result
FROM CRIMES
GROUP BY @\textcolor{red}{$A_{GB}$}@
HAVING result > @\textcolor{red}{\texttt{\$1}}@
\end{lstlisting}
\end{minipage}\\
\begin{minipage}{0.99\linewidth}
\captionsetup{singlelinecheck = false, justification=justified}
\lstset{upquote=true,frame=single, title = \bf{\QAJGH : join-groupby-having template }}
\begin{lstlisting}
SELECT @\textcolor{red}{$A_{GB}$,f($A_{agg}$)}@ AS result
FROM LINEITEM JOIN ORDERS ON L_ORDERKEY = O_ORDERKEY
GROUP BY @\textcolor{red}{$A_{GB}$}@
HAVING result > @\textcolor{red}{\texttt{\$1}}@
\end{lstlisting}
\end{minipage}\\
\begin{minipage}{0.99\linewidth}
\captionsetup{singlelinecheck = false, justification=justified}
\lstset{upquote=true,frame=single, title = \bf{\QAAJGH }}
\begin{lstlisting}
SELECT @\textcolor{red}{$A_{GB}$,f($A_{agg}$)}@ AS result
FROM (SELECT COUNT(*) AS COUNT_L, @\textcolor{red}{$A_{GB}$}@
      FROM LINEITEM GROUP BY @\textcolor{red}{$A_{GB}$}@)
      JOIN PART ON L_PARTKEY = P_PARTKEY
GROUP BY @\textcolor{red}{$A_{GB}$}@
HAVING result > @\textcolor{red}{\texttt{\$1}}@
\end{lstlisting}
\end{minipage}\\
\subsubsection{Samples}
We capture and store the samples through stratified sampling techniques based on the workload we introduced before. The stratified samples can be used for approximating result estimation and our cost estimation to help us to choose the optimal attribute.
\subsubsection{Attribute Selection Strategies}
\label{sec:strategies}
\BG{Currently only strategies that utilize single attributes, we need to motivate / justify this.}

 Many strategies can be applied to choose the attribute for building a provenance sketch for a query.  Different strategies include random picking from all attributes candidates after per-filtering (\stratRANall), randomly selecting from query-related attributes (\stratRANrelall), randomly selecting a group-by attribute (\stratRANgb), randomly selecting a primary key attribute (\stratRANpk). randomly selecting one of the aggregation input attribute  (\stratRANagg). These strategies utilizing random choice pick an attribute  uniformly at random from a set of candidate attributes specific to the strategy. For example, for randomly selecting a primary key attribute for a relation with a primary key $\{a,b\}$ we would pick each attribute with 50\% probability. In addition to the strategies that pick random attributes, we also evaluate strategies that use our size estimation techniques to select on attribute from a list of candidate attributes.
The \stratCBOPTrel strategy picks the attribute from the set of all safe attributes that is estimated to yield the smallest provenance sketch. The \stratCBOPTgb strategy applies the same selection criterion, but only considers group-by attributes.\BG{Why this strategy and not other subsets of attributes?}
To evaluate how our strategies compare against the optimal attribute choice (leading to the provenance sketch of minimal size), we also present results for this strategies which is based on our cost model to determine the optimal sketch (\stratCBOPT). What's more, \stratOPT is the actual optimal attribute and \stratNoPS is the original query without provenance sketch.

\subsection{Cost Estimation Accuracy \& Runtime}
In this section, we will talk about what our experiments measure. We run the auto-generated queries over \dsCrime, \dsTPCH and \dsParking, and applied our cost model to calculate the sketch data size. We compute the relative size error, accuracy for all queries. The number of range partitions of each provenance sketch is set 1000.
Firstly, We present the accuracy and effectiveness of our cost model based on the stratified sampling method. \BG{I would split this into two subsubsections: one for accuracy and one for ranking accuracy}
\subsubsection{Size Estimation Accuracy}\label{sec:size-est-accuracy}
In this subsection, Our initial evaluation focused on the disparity between the estimated and actual sketch data sizes. By running auto-generated queries on both \dsCrime, \dsTPCH and \dsParking datasets, we were able to gauge the accuracy of our estimation method. We measure the \emph{relative size error} \rse which is computed for a database $D$, query $\query$, and attribute $A$ using \aasize and \aesize as defined in section.\ref{section:rse}.
$$
\rse(\query,\db,A) = \frac{\absolute{\aesize - \aasize}}{\aasize}
$$

 Refer to Figures \ref{merged_relative_error} for a visual representation. For each dataset, \dsCrime, \dsTPCH, and \dsParking, we utilized the selection-aggregation-groupby template to randomly generate 1,000 queries. We present the results for two different sample rates: 5\% and 10\%. The plots clearly indicate that a higher sample rate leads to a reduced relative error. This can be attributed to the fact that a greater sample rate provides more tuples, offering richer information and thereby enhancing the accuracy of our estimation. The relative error for both \dsCrime and \dsParking is nearly zero, indicating that our estimations are accurate in most cases for these datasets. As for \dsTPCH, while the average relative error is slightly higher compared to \dsCrime and \dsParking, it remains within an acceptable range. 
From these plots, it's evident that the average error across each dataset is minimal which makes our cost model accurate and effective in choosing the optimal attribute.\\

    \subsubsection{Ranking Based on Estimation}
    \label{sec:rank-based-estim}
\begin{figure*}[htbp]
    \centering
    \begin{subfigure}[b]{0.45\textwidth}
        \centering
        \includegraphics[width=\textwidth]{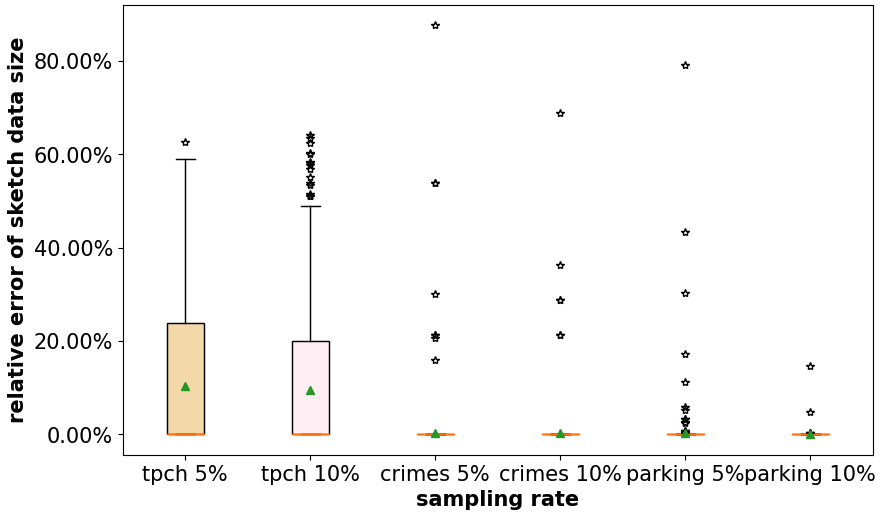}
        \caption{Relative sketch data size error varying sampling rate}
        \label{merged_relative_error}
    \end{subfigure}
    \hfill
    \begin{subfigure}[b]{0.45\textwidth}
        \centering
        \includegraphics[width=\textwidth]{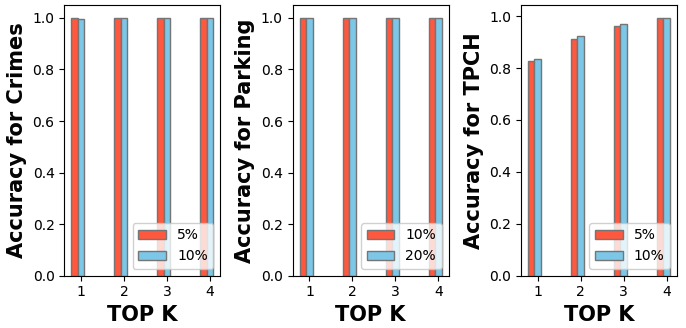}
        \caption{Accuracy of estimation of top k attribute varying k}
        \label{accuracy_merged_bar}
    \end{subfigure}
    \caption{Comparison of relative sketch data size error and accuracy estimation using \QAGH over 3 datasets.}
\end{figure*}

Absolute size estimates are important for determining whether it is worth to create a sketch, e.g., if a sketch is estimated to cover 80\% of the data then it is likely not useful for improving query performance. However, when selecting an attribute from a set of candidate attributes, the accuracy of size estimates is irrelevant as long as they are precise enough to rank the candidates correctly. Thus, we now evaluate the effectiveness of our estimation techniques for ranking candidate attributes.\BG{Merge all top-k related discussion from section above into here}
Thus, we measure the accuracy of ranking candidate attributes. Specifically, accuracy refers to the percentage of queries where the cost model's identified optimal attribute aligns with the actual optimal attribute within the TOP-K attributes. Solely relying on the top attribute can sometimes lead to missing the prime sketch candidate. To overcome this, our experiments encompassed the top-k attributes.  This approach, while potentially increasing overhead, aimed to strike a balance. By experimenting with different values of k, there is a trade-off between achieving greater accuracy and incurring higher overhead. This relationship is depicted in \Cref{accuracy_merged_bar}.
From the \Cref{accuracy_merged_bar}, For the \dsCrime and \dsParking dataset, our cost model consistently identifies the best attribute in nearly 100\% of the queries. In the case of \dsTPCH, selecting the top 2 attributes yields an 85\% accuracy rate, while considering the top 3 attributes boosts this to almost 100\%. These visual representations underline the efficacy of our cost model. For real-world scenarios, top-1 is ideal. In the case of TPC-H, we demonstrate that the top-1 sketch we chose is nearly optimal. The added expense doesn't justify the slight reduction in sketch size. The performance difference between \dsTPCH and the \dsCrime and \dsParking datasets can be attributed to the correlation among attributes. \dsCrime and \dsParking include geographical attributes that exhibit a higher correlation, whereas the attributes in \dsTPCH are less correlated. Datasets with a higher number of correlated attributes tend to perform better.


\subsection{Comparing Provenance Sketch Sizes for Our Strategies}
\label{different_strategies}
In this subsection we present comparative analysis of the sizes of provenance sketches generated using different strategies introduced in Section \ref{sec:strategies} across all three datasets. Our approach involves computing the relative sketch data size, Expected size and runtime varying strategies on the queries from \QAGH template and \QAJGH template workload. These templates are safe for our Provenance Sketch and encompass most of our routine use cases.

\subsubsection{Relative sketch data size of random strategies}
The relative sketch data size represents the proportion of the provenance sketch size for each candidate attribute compared to the total dataset size.This evaluation is illustrated in \Cref{AVG_sketch_selectity} providing a comprehensive understanding of the merits of each strategy. From the results, we can see for all three datasets, the relative sketch data size for \stratCBOPT is close to the \stratOPT, demonstrating that our cost model is effective for choosing the optimal one.\\
\subsubsection{Expected size of random strategies}
To be able to compare strategies that select one attribute based on size estimations with strategies that randomly select an attribute from a set of candidate attributes, we report the expected size of random strategies. As all of these strategies pick an attribute $\att$ from a set of candidate attributes $\candA$ uniformly at random, the expectation is equal to the average sketch size across all attributes from $\candA$:
\begin{align*}
\expectation{\aasize}{\att} &= \frac{\sum_{\att \in \candA} \aasize}{\card{\candA}}
\end{align*}
It's important to note that when it comes to strategies based on random selection, we operate under the assumption of a uniform distribution, ensuring consistency in our approach. The results are shown in \Cref{box_plot_merged}. We can conclude from plot that the expected size of \stratCBOPT and
\stratOPT are quite close, proving the effectiveness of our cost model. \stratRANgb has the minimal expected size comparing to \stratRANagg and \stratRANrelall

 \subsubsection{Average runtime}
\begin{figure*}[htbp]
    \centering
    \begin{subfigure}[b]{0.32\textwidth}
        \centering
        \includegraphics[width=\textwidth]{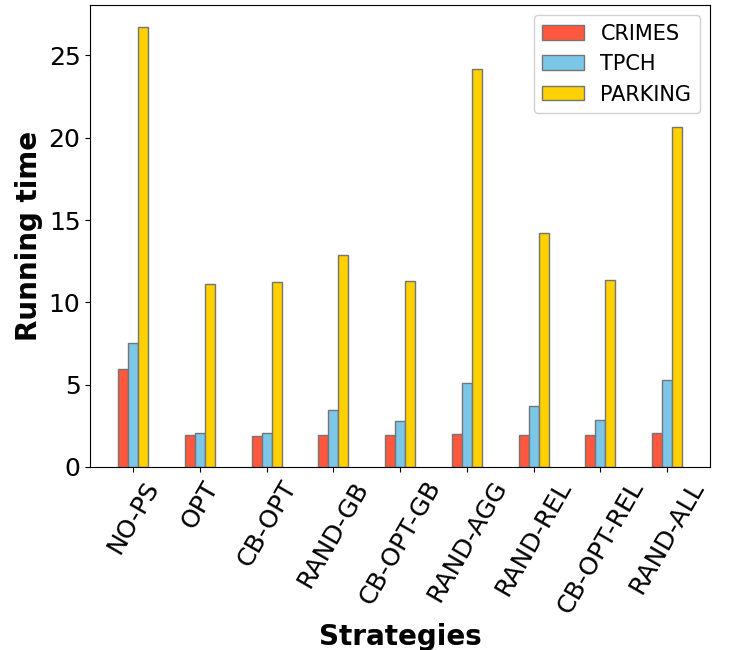}
        \caption{Average runtime of queries using sketches}
        \label{running_time_comparation}
    \end{subfigure}
    \hfill
    \begin{subfigure}[b]{0.30\textwidth}
        \centering
        \includegraphics[width=\textwidth]{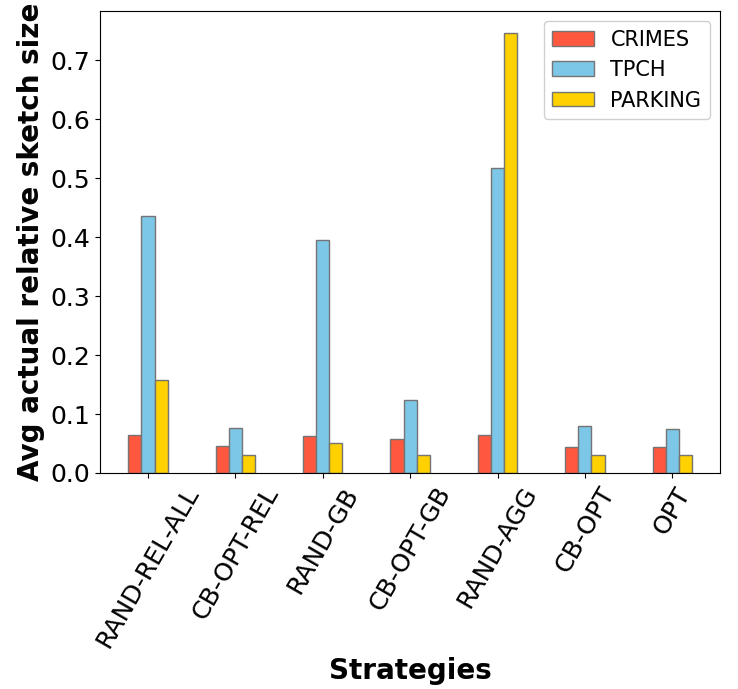}
        \caption{Average relative sketch data size}
        \label{AVG_sketch_selectity}
    \end{subfigure}
    \hfill
    \begin{subfigure}[b]{0.34\textwidth}
        \centering
        \includegraphics[width=\textwidth]{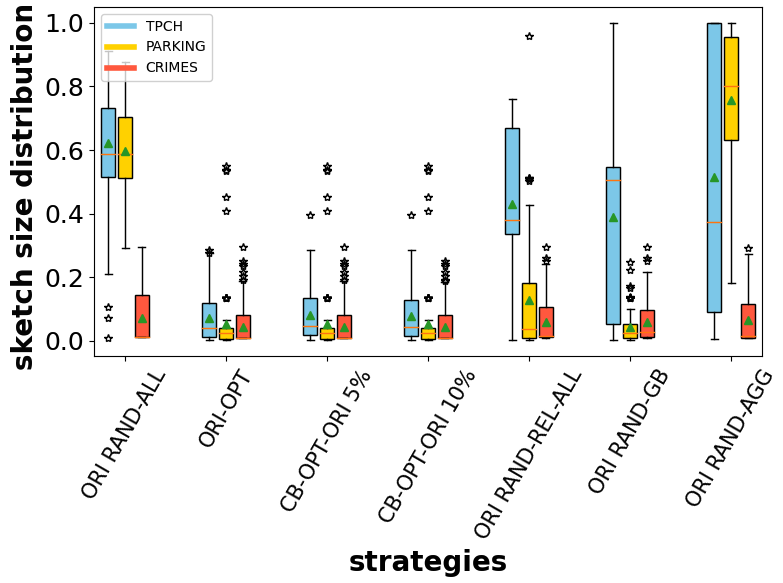}
        \caption{Distribution of relative sketch data size for 1000 queries}
        \label{box_plot_merged}
    \end{subfigure}
    \caption{Comparison of runtime, average relative sketch size, and distribution across 3 datasets using \QAGH}
\end{figure*}

Apart from the relative sketch data size, we also examined the average running time of queries for different strategies. The workload is same as the workload of average relative sketch data size test. In this test, we captured, used and ran queries with provenance sketches based on the selection of various strategies. We observed that the results were consistent with those obtained from the relative sketch data size analysis. To effectively illustrate these findings, we plot the results using \Cref{running_time_comparation}. From the results, average runtime in \stratCBOPT is a little bit higher than \stratOPT.   

 \subsubsection{experiments conclusion}
 From the experiments performance before. for the random strategies. \stratRANgb has the best performance. We can see the performance of \stratCBOPTgb is better than \stratRANgb. the performance of \stratCBOPTgb and \stratCBOPTrel is close to the \stratOPT, a bit higher than the \stratOPT. However, \stratCBOPTrel requires more overhead cost because, the relative candidate is more than the group by attribute. As a result, the \stratCBOPTgb is a better choice.

\subsection{End-to-end}\label{sec:experiments-end-to-end}

We now measure the end-to-end runtime of evaluating full workloads of \QAGH and \QAJGH templates. For each strategy we start from an empty set of provenance sketches. The purpose of these experiments is to provide a holistic view of the performance of our strategies in a realistic setting, encompassing capturing provenance sketches (PS), sorting the data\BG{What for, needs explanation} \textbf{as we discussed in \cref{section:samplings}}, sampling, and ultimately executing queries on the PS. For each incoming query, if there is a provenance sketch that can be utilized then we instrument the query to use the sketch. Otherwise, we create a new sketch using the strategy to decide what sketch to create.
If the data is sampled and stored before, we directly operates size estimation and choose the optimal candidate in the strategy. Otherwise, the stratified samples should be captured first. Then stored them for next time reuse. The next step is capturing the provenance sketch and store this provenance sketch for reusing. After capturing the PS, we transition to measuring the speed of running queries on it.

\begin{figure*}[htbp]
    \centering
    \begin{subfigure}[b]{0.24\textwidth}
        \centering
        \includegraphics[width=\textwidth]{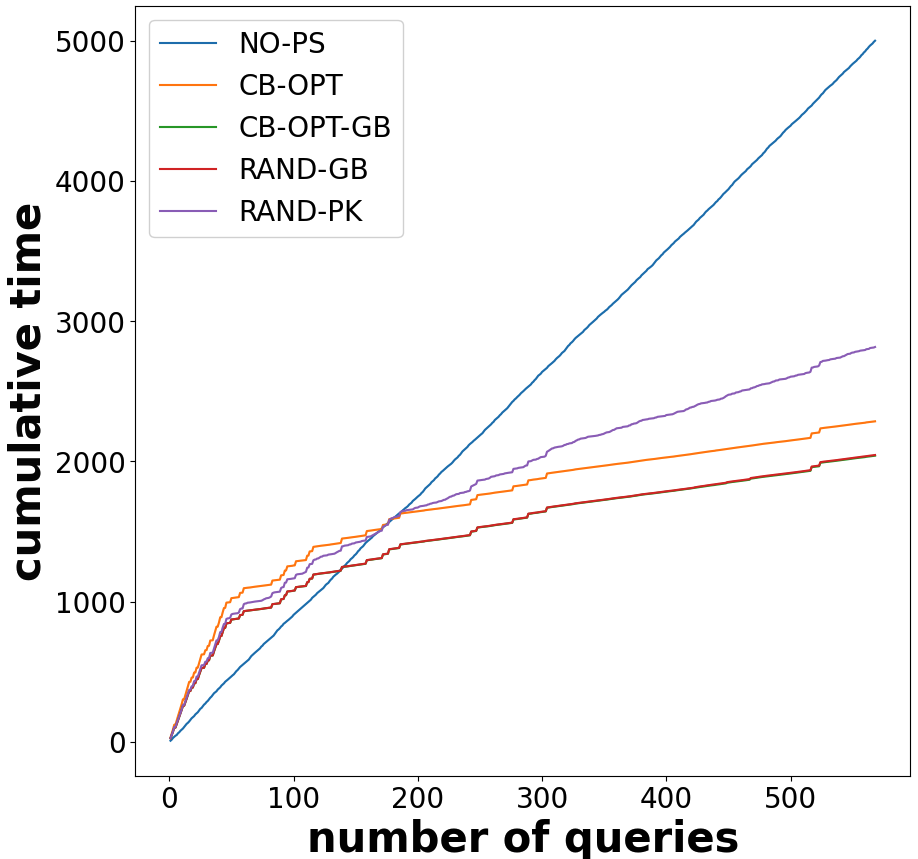}
        \caption{2 group by over \dsTPCH}
        \label{end-to-end-different-strategies}
    \end{subfigure}
    \hfill
    \begin{subfigure}[b]{0.24\textwidth}
        \centering
        \includegraphics[width=\textwidth]{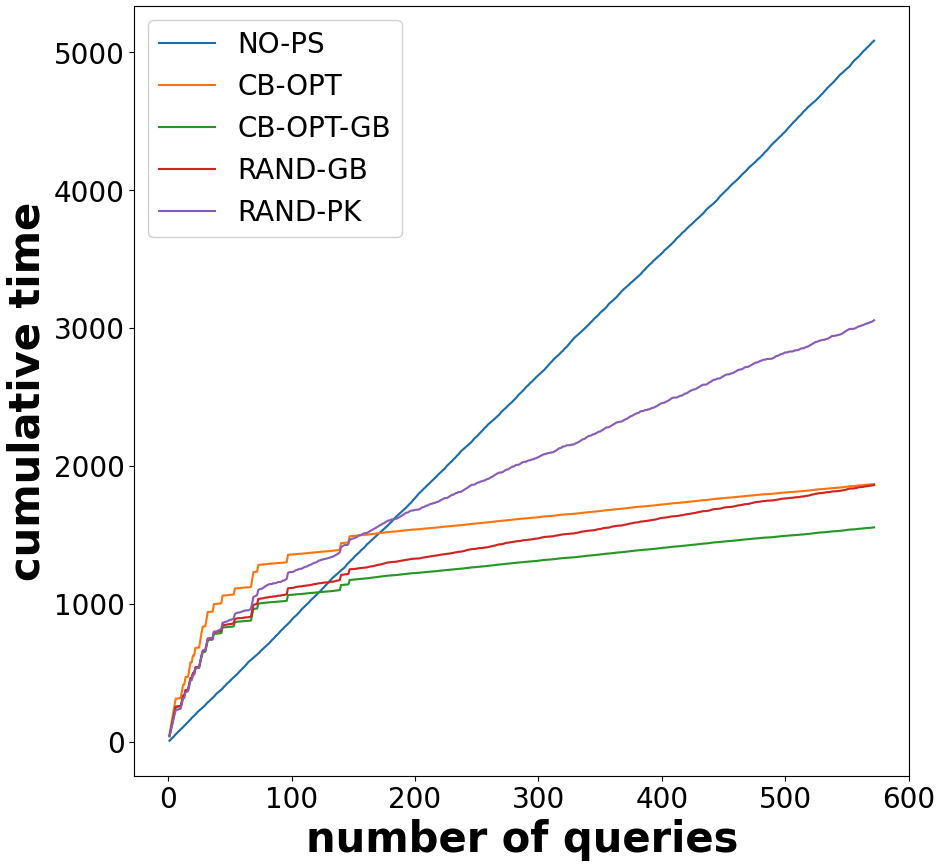}
        \caption{3 group by over \dsTPCH}
        \label{end-to-end-different-strategies-3gb}
    \end{subfigure}
    \hfill
    \begin{subfigure}[b]{0.24\textwidth}
        \centering
        \includegraphics[width=\textwidth]{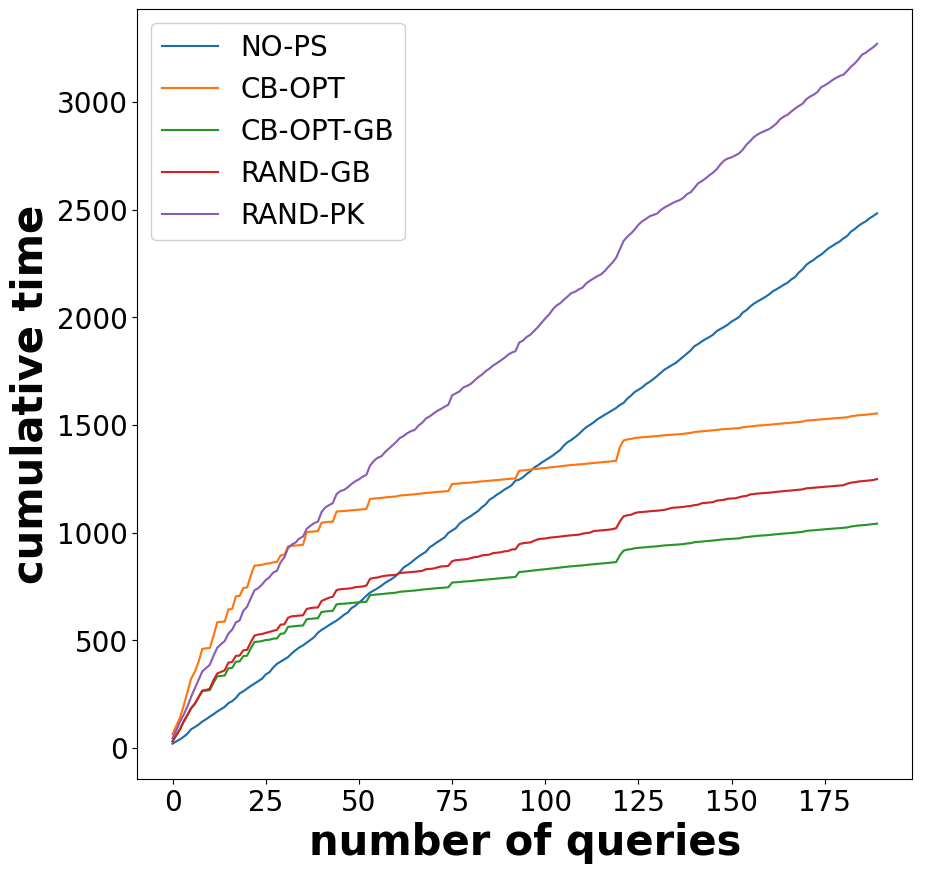}
        \caption{2 group by over \dsTPCH (join)}
        \label{end-to-end-different-strategie-wdj}
    \end{subfigure}
    \hfill
    \begin{subfigure}[b]{0.24\textwidth}
        \centering
        \includegraphics[width=\textwidth]{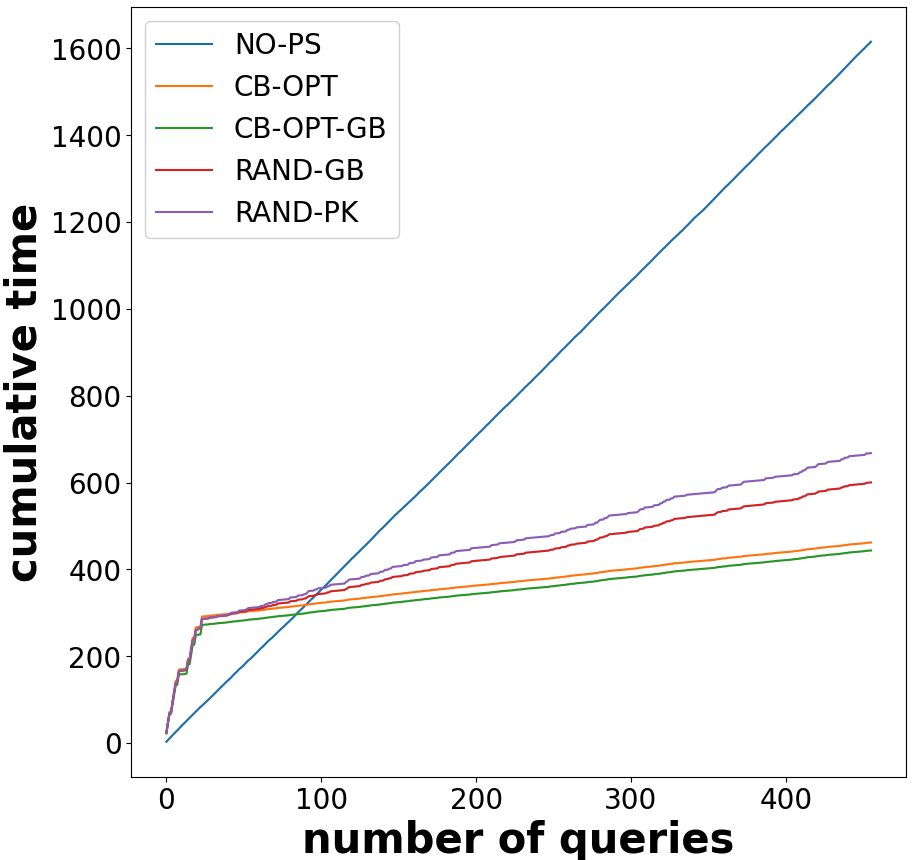}
        \caption{3 group by over stars}
        \label{end-to-end-different-strategies-3gb-stars}
    \end{subfigure}
    \caption{Cumulative time comparison of different strategies for selection-aggregation queries over \dsTPCH and stars datasets.}
\end{figure*}

The empirical results derived from these experiments are lucidly shown in  \Cref{end-to-end-different-strategies,end-to-end-different-strategies-3gb}, \ref{end-to-end-different-strategie-wdj} and \ref{end-to-end-different-strategies-3gb-stars}. Analyzing these plots, several conclusions could be made. The \stratCBOPTgb strategy consistently has the best performance comparing the competitors, \stratRANgb have the second best performance, whereas \stratRANpk has the worse performance comparing the former 2 strategies. From the results, in the beginning, the workload runtime cost is higher than the original workload without provenance sketch. That's we need pay overhead cost, including sampling, estimating and choosing the optimal one. After enough provenance sketches has been create and stored, no new provenance sketch are needed to be created, the upcoming queries can directly reuse the provenance sketch which have been created before. We can benefit from our end-to-end workload.



\subsection{Summary}
In our experiments section. We test the accuracy of our cost model, showing the effectiveness of our cost model to determine the optimal attribute for capturing provenance sketch. THE expected size and runtime experiments indicate that \stratCBOPTgb has the best performance comparing with the other strategies in the aspect of runtime and selectivity. The end-to-end experiments shows that we can benefit from our cost model to determine the optimal attribute for capturing the provenance sketch for a large workload with the common use case queries, even though the upfront cost needs to be paid.

\section{Conclusions and Future Work}
\label{sec:concl-future-work}
We present cost model for provenance-based data skipping (PBDS), a technique that determine which attribute is the optimal one for capturing provenance sketch. We combine the stratified sampling and approximating query processing and develop the cost model rules. The cost model results in good performance for estimating cost and optimizing the capturing process of PBDS for the important classes of queries. In the future, we will investigate how to extend our cost model for maintaining provenance sketch. What's more, we will study in the relationship of relative attributes'correlation in query with the cost of the provenance sketch.

%
%



\bibliographystyle{ACM-Reference-Format}
\bibliography{2022_Cost_Model_For_Provenance_Sketches-main}

\end{document}